\documentclass[usenatbib]{mnras}

\usepackage[T1]{fontenc}
\usepackage{ae,aecompl}

\usepackage{comment}
\usepackage{graphicx}	
\usepackage{color}
\usepackage{journals}
\usepackage{multirow}
\usepackage{rotating}
\usepackage{xcolor,listings}
\usepackage{textcomp}
\usepackage{etoolbox}
\makeatletter
\patchcmd\@combinedblfloats{\box\@outputbox}{\unvbox\@outputbox}{}{%
  \errmessage{\noexpand\@combinedblfloats could not be patched}%
}%
\makeatother

\newcommand{\Massunh}{$\rm h^{-1}{\cal M}_\odot$}

\definecolor{grey}{rgb}{0.4,0.6,0.6}
\definecolor{brown}{rgb}{0.65,0.16,0.16}
\definecolor{darkgreen}{rgb}{0.0,0.7,0.0}

\usepackage{amssymb}
\usepackage{pifont}
\newcommand{\xmark}{\ding{53}}%

\begin{document}

\title [Frequency and nature of compact groups from SAMs]
{Compact groups from semi-analytical models of galaxy formation -- I:
   a comparative study of frequency and nature
  }
\author[D\'iaz-Gim\'enez et al.]
{E. D\'iaz-Gim\'enez$^{1,2}$\thanks{eugeniadiazz@gmail.com}, 
A. Taverna$^{1,2}$,
A. Zandivarez$^{1,2}$,
G. A. Mamon$^{3}$ 
\\
\\
$1$ Universidad Nacional de C\'ordoba (UNC). Observatorio Astron\'omico de C\'ordoba (OAC). C\'ordoba, Argentina\\
$2$ CONICET. Instituto de Astronom\'ia Te\'orica y Experimental (IATE). C\'ordoba, Argentina \\
$3$ Institut d'Astrophysique de Paris (UMR 7095: CNRS \& Sorbonne Universit\'e), Paris, France}
\date{\today}
\pagerange{\pageref{firstpage}--\pageref{lastpage}}
\maketitle
\label{firstpage}

\begin{abstract}

Compact groups (CGs) of galaxies are defined as isolated and dense galaxy systems that appear to be a unique site of multiple galaxy interactions. Semi-analytical models of galaxy formation (SAMs) are a prime tool to understand CGs. We investigate how the frequency and the three-dimensional nature of CGs depends on the SAM and its underlying cosmological parameters. Extracting 9 lightcones of galaxies from 5 different SAMs and selecting CGs as in observed samples, we  find  that  the frequency and nature of CGs depends strongly on the cosmological parameters. Moving from the WMAP1 to the WMAP7 and Planck cosmologies (increasing density of the Universe and decreasing normalisation of the power spectrum), the space density of CGs is decreased by a factor 2.5, while the fraction of CGs that are physically  dense  falls  from  50  to  35 percent. The  lower $\sigma_8$
leads  to  fewer  dense  groups, while the higher $\Omega_{\rm m}$ causes more chance alignments. However, with increased mass and spatial resolution, the fraction of CGs that are physically dense is pushed back up to 50 percent. The intrinsic differences in the SAM recipes also lead to differences in the frequency and nature of CGs, particularly those related to how SAMs treat orphan galaxies. We find no dependence of CG properties on the flux limit of the mock catalogues nor on the waveband in which galaxies are selected. One should thus be cautious when
interpreting a particular SAM for the frequency and nature of CGs.
\end{abstract}



\section{Introduction} 
Compact groups (CGs) of galaxies represent an extreme galaxy environment where the space density of galaxies appears to be as high as the cores of rich clusters, yet the  CGs are designed to be isolated. The high space densities and low velocity dispersions of CGs make them the ideal sites for galaxy mergers \citep{Mamon92} and rapid interactions. 

Since the first studies of  CGs by \cite{Stephan1877}, 
\cite{Seyfert48}, \cite{Shakhbazyan73},
and \cite{Robinson&Wampler73}, the study of these systems became an important
research field in extragalactic astronomy. The statistical properties of galaxies in CGs have been determined thanks to the construction of CG catalogues
(e.g. \citealt{Shakhbazyan73,Petrosian74,Rose77}), and especially the popular
Hickson Compact Group catalogue (HCG, \citealt{Hickson82,Hickson93},
and references therein). Several subsequent attempts were devoted to the
construction of catalogues via automatic search algorithms
(e.g. \citealt{Mamon89} for a search within the Virgo cluster yielding a
single group;
\citealt{Prandoni+94,Focardi&Kelm02,Lee+04,McConnachie+09,DiazGimenez+12,Hernandez&Mendes15,sohn+15,sohn+16,DiazGimenez+18}).

The CG environment indeed appears special given the high frequency of interacting galaxies within them \citep{MendesdeOliveira&Hickson94}. However, if CGs are as dense as they appear in projection on the sky, their fractions of spiral morphologies are much higher \citep{Mamon86} suggesting either recent formation \citep{Hickson&Rood88} or pollution of the HCG sample by CGs caused by chance alignments of galaxies along the line of sight \cite{Mamon86}. In fact, chance alignments of galaxies within much larger typical groups of galaxies are frequent enough to roughly predict the frequency of isolated compact configurations satisfying HCG criteria \citep{Mamon86,Walke&Mamon89}. Moreover, the nearest CG satisfying the HCG criteria, found in the automated search in the Virgo cluster (\citealp{Mamon89}, around M60), is sufficiently close that redshift-independent distances are precise enough to infer its three-dimensional nature, and it turns out to be a chance alignment \citep{Mamon08}.

One can resort to cosmological simulations to infer the nature and frequency of CGs. Several teams have extracted CGs using \emph{semi-analytic models of galaxy formation and evolution} (SAMs) to this end.
SAMs employ physical recipes to describe galaxies, usually associated with the subhaloes of the haloes of a previously run cosmological simulation (without gas). Given the positions of galaxies at different timesteps, one can build a lightcone to derive a mock observational sample of galaxies in redshift space (sky position and redshift). One can then run an algorithm to extract CGs from this lightcone.

In recent years, numerous works have embraced the task of comparing results
from different SAMs applied on the same simulations, looking for similarities and/or differences.
For instance,
\cite{Maccio+10} studied the radial distribution and luminosity function
(LF) of Milky Way satellites using 3 SAMs.
\cite{Dariush+10} and later \cite{snaith11} studied the difference between
first and second most luminous galaxies in groups using 2 or more SAMs.
\cite{DiazGimenez&Mamon10} (hereafter DGM10) analysed the properties of compact groups extracted from three different SAMs, all of them
applied on the Millennium I Simulation \citep{Springel+05}. DGM10 found
that the fraction of CGs that were not truly dense in real space
varied from 1/4 to 2/5 
depending on the SAM.
Using the same simulation, \cite{delucia10} found that different treatments
of galaxy mergers in  SAMs led to  different merger time scales, with important implications for the formation and evolution of massive galaxies. 
More recently, \cite{lu14} used the merger trees extracted from the Bolshoi simulation \citep{klypin11} as input of three SAMs. They concluded that, in spite of the different parametrisations of star formation and feedback processes, the three models yielded similar qualitative predictions for the evolutionary history of galaxy stellar masses and star formation.
\cite{Contreras13} analysed the two point correlation function of galaxies in
two different SAMs, concluding that the 1-halo term (pairs in the
same dark matter halo) is sensitive to the subhalo finder, which affects the radial
distribution of galaxies.
\cite{gozaliasl16,gozaliasl18} studied the evolution of the brightest group galaxy stellar mass, star formation rate and its contribution to the total baryonic mass of the halos in four different SAMs run on the Millennium I simulation and compared their predictions with observations. They found 
that SAMs predictions show difficulties to reproduce the evolution of the brightest group galaxy stellar mass as well as the fraction of baryonic mass for $z\ge 0.6$.  

The study by \cite{knebe15} was the first of a series comparing 12 different
SAMs as well as 2 halo occupation models.
They found that, without calibration, the 
14 models show important scatter in stellar mass functions, stellar-halo mass
relations, and specific star formation rates, which they show
to originate to  the difference in cosmological parameters and the different
treatments of orphan galaxies.
\cite{Asquith+18} extended this study to higher redshifts and to a comparison
with observations, to conclude that many models require a physical recipe
dissociating the growth of low mass galaxies from that of their dark matter haloes.
Finally, \cite{pujol17} found that the clustering of galaxies depends on the treatment
of orphan galaxies.

Several studies used SAMs to understand the nature and frequency of CGs. \cite{McConnachie+08} found that 30 per cent of CGs selected in projection on the sky (before using redshift information) are physically dense in three dimensions (3D).
As mentioned above, DGM10 found that from 3/5 to 3/4 of CGs selected after redshift filtering of obvious interlopers are physically dense, depending on which of the 3 SAMs they analysed. This was the first study to consider galaxies as extended instead of point masses and consider possible blending of nearby galaxies close aligned along the line of sight. 
Moreover, DGM10 found that the frequency (space density) of CGs was roughly 10 times higher in the simulations than observed, suggesting that the HCG sample was 90 per cent incomplete, mainly in groups dominated by a single galaxy (as previously noted by \citealp{Prandoni+94}). In subsequent studies by our team \citep{DiazGimenez+12,DiazGimenez+15,Taverna+16,DiazGimenez+18}, CGs were extracted from the public outputs of different SAMs based on different physics or from those of older SAMs run on the dark matter haloes extracted from cosmological simulations run with different cosmological parameters. These studies
led to different fractions of physically dense groups.

This variation in the fractions of physically dense CGs on one hand and of the frequency of CGs on the other has motivated us to perform a thorough study on how the SAM and its parent cosmological simulation affect our interpretations of CGs.

In this work, we use several publicly available SAMs that have not been tailored to specifically reproduce physical properties of extreme galaxy systems such as CGs. We test the performance of the different SAMs run on different cosmological simulations when identifying CGs. The majority of the models used here are part of the Millennium run simulation project \citep{virgo}. To improve our comparison, we have also included a SAM that was implemented in a different large cosmological simulation performed by the MultiDark project\footnote{https://projects.ift.uam-csic.es/multidark/} \citep{knebe+18}. 
 
The layout of this work is as follows. We present the SAMs in Sect.~\ref{sec:simulations}. In Sect.~\ref{sec:lightcones}, we explain how we built the mock lightcones and their main properties, while in Sect.~\ref{sec:sample} we describe the construction of the CG samples. Finally, we discuss our results in Sect.~\ref{sec:results} and summarise our conclusions in Sect.~\ref{sec:conclusions}.

\section{Semi-analytical models of galaxy formation}
\label{sec:simulations}

\begin{table*}
\begin{center} 
\caption{Semi-analytical models studied in this work. \label{tab:sams} }
\tabcolsep 2pt
\begin{tabular}{lllcccclccccc}
\hline
\hline
\multicolumn{1}{c}{Author} & acronym & \multicolumn{4}{c}{Cosmology} & & \multicolumn{4}{c}{Simulation} & & Limits \\
\cline{3-6} \cline{8-11} 
			&& \multicolumn{1}{c}{name} & $\Omega_{\rm m}$ & $h$ & $\sigma_8$ & &
\multicolumn{1}{c}{name} & box size & $\epsilon$ & $\log\,m_{\rm p}$ & & $\log\,{\cal M}^*$ \\
			&&  	&  	&      &     & & & $[h^{-1} \,\rm Mpc]$ &$[h^{-1} \,\rm kpc]$ &
$[h^{-1}\,{\cal M}_\odot]$& & $[$\Massunh$]$ \\
\multicolumn{1}{c}{(1)} & \multicolumn{1}{c}{(2)} & \multicolumn{1}{c}{(3)} & (4) & (5) & (6) & & \multicolumn{1}{c}{(7)} & (8) & (9) & (10) && (11) \\
\hline
\cite{DeLucia+Blaizot07}& DLB$^{\rm b}$&WMAP1 	   &0.25& 0.73 & 0.90 &&
MS& \ 500 & 5 & 8.94 && 8.85  \\
\cite{Guo+11}           & G11$^{\rm b}$&WMAP1 	   &0.25& 0.73 & 0.90 &&
MS& \ 500 & 5 & 8.94  && 8.85  \\
\cite{Guo+11}  		& GII$^{\rm b}$&WMAP1 	   &0.25& 0.73 & 0.90 &&
MSII& \ 100 & 1 & 6.84  && 7.85 \\
\cite{Guo+13}       & G13$^{\rm b}$&WMAP7 	   &0.27& 0.70 & 0.81 &&
MS& \ 500  & 5 & 8.97 && 8.85  \\
\cite{Henriques+15}$^{*}$   & HrI$^{\rm b}$&Planck&0.31& 0.67 & 0.83 &&
MS& \ 480  & 5 & 8.98 && 8.85  \\
\cite{Henriques+15}$^{*}$   & HrII$^{\rm b}$&Planck&0.31& 0.67 & 0.83 &&
MSII& \ \ 96 & 1 & 6.89  && 7.85   \\
\cite{Cora+18}		& SAG$^{\rm b}$&Planck	&0.31	& 0.68 & 0.82 && 
MDPL2& 1000 & 13$\,\to\,$5 & 9.18 && 8.85  \\
\hline
\end{tabular}  
\end{center}
\parbox{\hsize}{\noindent Notes: The columns are:
(1): authors;
(2): acronym of SAM;
(3): cosmology of parent simulation;
(4): density parameter of parent simulation;
(5): dimensionless $z$=0 Hubble constant of parent simulation;
(6): standard deviation of the (linearly extrapolated to $z$=0) power spectrum on the scale of $8\,h^{-1}\,\rm Mpc$;
(7): name of parent simulation (MS = Millennium I, MSII = Millennium II, MDPL2 = MultiDark Planck 2);
(8): periodic box size of parent simulation;
(9): force softening of parent simulation (the force softening of MDPL2 varies from $13\,h^{-1}\,\rm kpc$ at high redshift to $5\,h^{-1}\,
\rm kpc$ at low redshift);
(10):  log particle mass of the parent simulation;
(11):  lower limit of stellar mass of galaxies.
The `b' superscript in the acronyms stands for `box' (to
  distinguish from lightcones built in Sect.~\ref{sec:lightcones}).
  More accurate values for the simulations can be found at
\url{http://gavo.mpa-garching.mpg.de/Millennium/Help/simulation} and \url{https://www.cosmosim.org/cms/simulations/mdpl2/}.}
\\
\parbox{\hsize}{\noindent  $^{*}$ HrI and HrII SAMs were run on re-scaled
  versions of the original Millennium simulations (hence, the different box
  sizes and particle masses).}
\end{table*}

SAMs encompass the main physical processes
that govern the formation and evolution of galaxies in a set of parameterized, 
self-consistent and iterative differential equations. %
Some of these processes are gas infall and cooling, reionization of the
Universe, star formation, growth of the central black hole, feedback by
active galactic nuclei (AGN) and supernovae (chemical enrichment), mergers of galaxies, 
photometric evolution, among others. 
In practice, these equations describe how the baryons move between different reservoirs of mass (haloes of dark matter).
Through an analytical treatment, these baryons are related to the merger trees and are followed back in time. 
The parameters of the models are tuned in order to obtain results that approximate in the best possible way the main
observational properties of the galaxy population of the local Universe, such as colour, luminosity and/or stellar mass functions, or even at higher redshifts.

The galaxy properties that SAMs try to reproduce 
may be affected by the physical recipes used in these models.
For example, in SAMs, feedback from AGN  reduces the 
luminosity and stellar mass of the brightest galaxies, 
while supernovae are more effective in removing gas from  low mass galaxies, 
thus decreasing their star formation rate. 
By taking into account AGN, supernovae and other physical processes,
SAMs have succeeded in reproducing many important observable properties of galaxies. 
However, these physical processes were treated in different ways by different authors, 
resulting in different solutions to the problem of galaxy formation, and the
agreement with observations may be fortuitous as the modelling of physical
recipes is probably over-simplistic in all SAMs.

We analyzed four SAMs run on the Millennium simulation (MS,
\citealp{Springel+05}), two SAMs run on the Millennium simulation II (MSII,
\citealp{BoylanKolchin+09}), and one SAM run on the Multidark Simulation
(MDPL2, \citealp{knebe+18}). These SAMs are those of:
\vspace{-0.5\baselineskip}
\begin{itemize}
    \item \cite{DeLucia+Blaizot07} run on the original MS with its WMAP1 cosmology  \citep{Spergel+03};
    \item \cite{Guo+11} run on the original MS;
    \item \cite{Guo+11} run on the higher resolution (and smaller box) MSII;
    \item \cite{Guo+13} run on the MS 
      with a WMAP7 cosmology \citep{Komatsu+11};
    \item \citeauthor{Henriques+15} (\citeyear{Henriques+15}, L-Galaxies), run on the MS re-scaled to the Planck cosmology \citep{Planck+16};
    \item \cite{Henriques+15} run on the MSII re-scaled to the Planck cosmology;
    \item \citeauthor{Cora+18} (\citeyear{Cora+18}, SAG), run on the MDPL2 with Planck cosmology.
\end{itemize}
 In Appendix~\ref{queries}, we quote the queries used to retrieve data from the public outputs of the SAMs. Following \cite{Guo+11,knebe15,irodotou19}, we retrieved galaxies from the databases with stellar masses larger than $\sim 10^9 \, {\cal M}_\odot$ ($7\times 10^8 \, h^{-1} {\cal M}_\odot$ )
 for the MS- and MDPL2-based SAMs, 
 and stellar masses largen than $\sim 10^8 \, {\cal M}_\odot$ ($7\times 10^7 \, h^{-1} {\cal M}_\odot$) 
 for the MSII-based SAMs.
 In Table~\ref{tab:sams}, we list the different SAMs with their corresponding
parent simulations and cosmological parameters.

\begin{figure}
  \centering
  \includegraphics[width=\hsize,viewport=0 30 400 550]{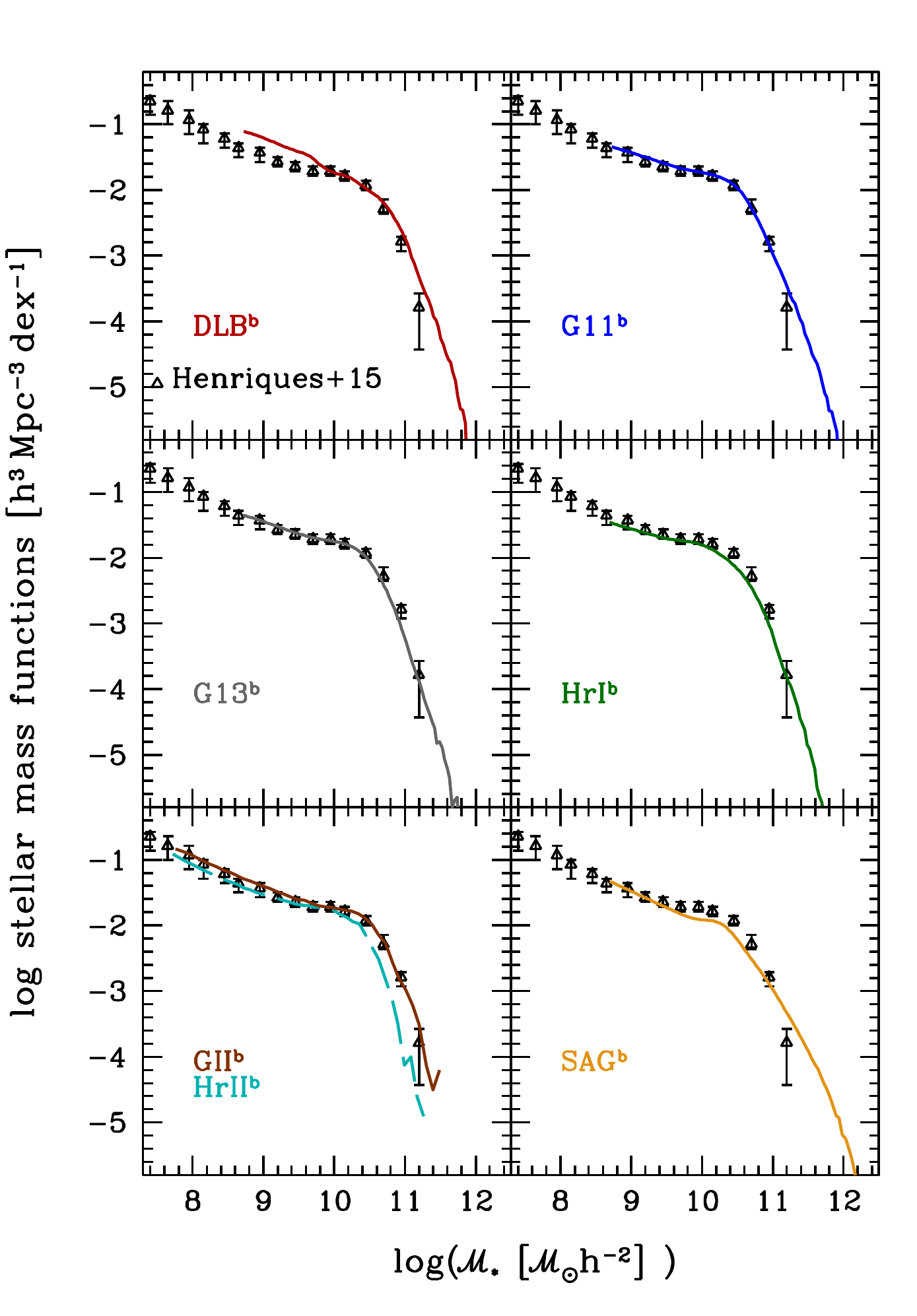}
  \caption{
Galaxy stellar mass functions in the simulation box at $z=0$ for
different SAMs (\emph{curves}) compared with the observational computation of
Henriques et al. (2015)
for the combined observational data from SDSS
(Baldry et al. 2008; Li et al. 2009)
and GAMA
Baldry et al. (2012)
catalogues (\emph{symbols}).
  }
\label{fig:smf}
\end{figure}

We now illustrate the differences in outputs of the 7 SAMs according to their
stellar mass functions (SMF),
luminosity functions (LF), 
morphological fractions vs. stellar mass, 
and 2-point correlation functions, all measured in their last snapshot, roughly corresponding to $z=0$.

Figure~\ref{fig:smf} shows the stellar mass functions (SMF) of the 7 SAMs
compared to the observations from the SDSS and Galaxy and Mass Assembly
(GAMA) surveys compiled by \cite{Henriques+15}.  From these comparisons, we see that
DLB$^{\rm b}$ overproduces galaxies with low and high stellar masses.
G11$^{\rm b}$ used the SMF s calibration of their SAM,
and therefore they obtained a better agreement, although it still
overproduces galaxies at the highest end of the SMF. 
G13$^{\rm b}$ achieve a very good agreement with the observations 
in the whole range of stellar masses, 
while HrI$^{\rm b}$ shows a slightly lower density of galaxies near
the characteristic stellar mass.  In the Millennium II, GII$^{\rm b}$
performs really well in the whole range of masses, while HrII$^{\rm b}$
underproduces galaxies at the highest end of the SMF, but works well for
masses lower than $\sim 10^{10.5} {\cal M}_\odot h^{-2}$. 
Lastly, SAG$^{\rm b}$ 
overproduces galaxies at the highest end of the SMF while it lacks galaxies at the knee of the SMF. 
All SAMs but DLB have used the observational SMF as calibration of their SAMs.
HrI, HrII, SAG have used the observations shown in this figure, 
while G11 and G13 have used some of the data that conform the compilation of observations performed by \cite{Henriques+15}.

\begin{figure}
\centering
\includegraphics[width=\hsize,viewport=0 30 400 550]{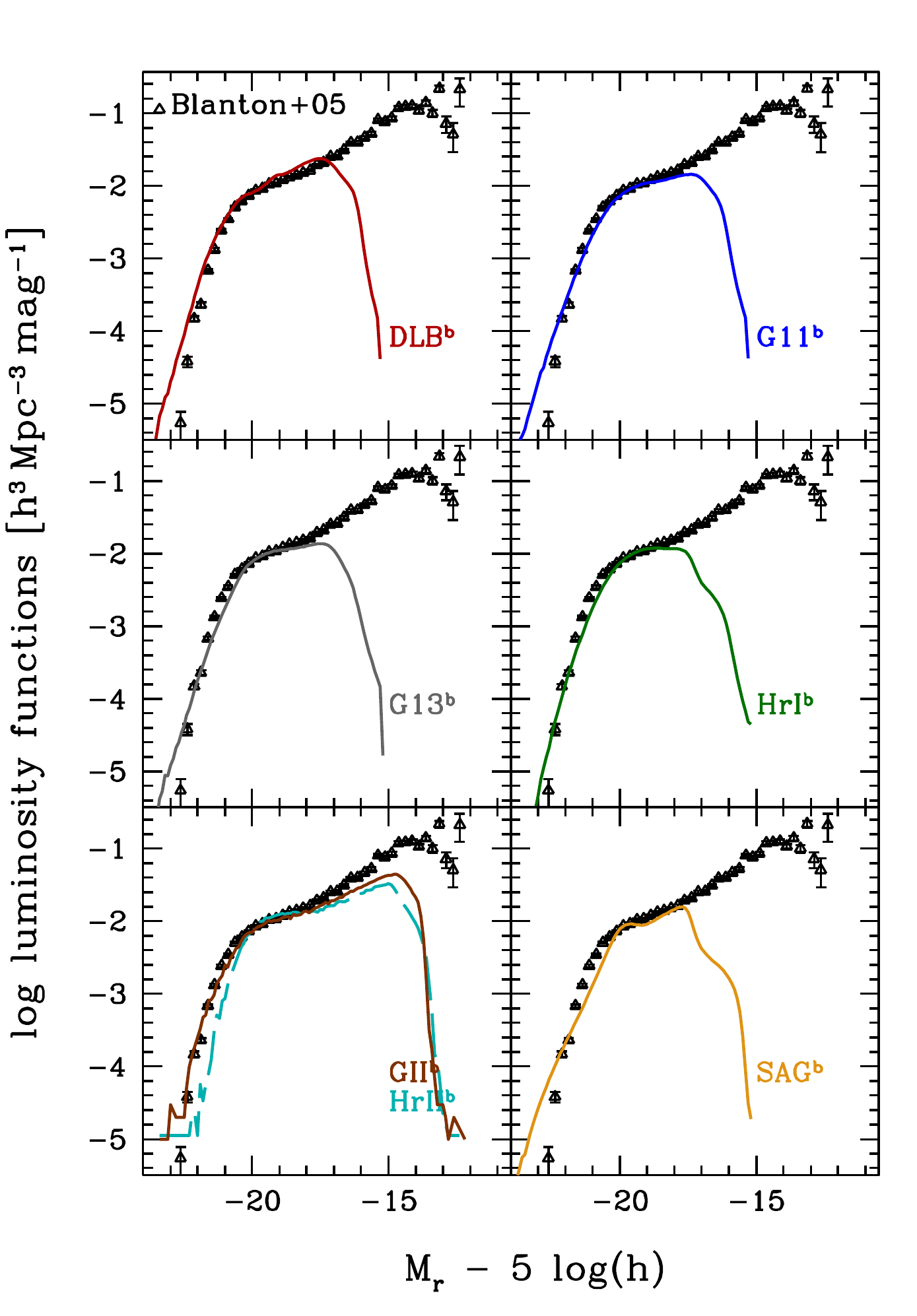}
\caption{Luminosity functions of galaxies in the simulation box at
  $z=0$ in the SDSS $r$-band for different SAMs (\emph{curves}, acronyms given in
  Table~\ref{tab:sams}) and extracted from the SDSS DR2 observations by
Blanton et al. (2005)
(\emph{symbols}).
}
\label{fig:fdel}
\end{figure}
\nocite{Blanton+05}
\nocite{Henriques+15}
\nocite{baldry08}
\nocite{li09}
\nocite{baldry12}

Figure~\ref{fig:fdel} compares the $z$=0 $r$-band luminosity functions for the 7 SAMs to the observational
data by \cite{Blanton+05} using the Sloan Digital Sky Survey Data Release 2
(SDSS DR2, \citealt{sdssdr2}). Since we have chosen to work with samples that are complete in stellar masses, the LFs of these galaxies in the boxes show incompleteness at low luminosities, being complete up to $M_r - 5 \log h \sim -17$ for MS-based SAMs and SAG, and $M_r - 5 \log h \sim -15$ for the two MSII-based SAMs (GII$^{\rm b}$ and HrII$^{\rm b}$). 
Relative to the observations, the SAMs
generally reproduce reasonably well the LF.
In particular, DLB$^{\rm b}$ overpredicts both the bright end of the LF and the range of luminosities above the knee ($M^{\ast}$).
Meanwhile, G11$^{\rm b}$, G13$^{\rm b}$ and HrI$^{\rm b}$ match well the
observed LF around $M^{\ast}$ and brighter, but
underpredict the LF at low luminosities ($M_r \sim -18$).
GII$^{\rm b}$ shows a good match to the observed LF for galaxies more luminous than $M_r = -17$, while HrII$^{\rm b}$
underpredicts the bright end of the LF. Although these SAMs show a lack of
galaxies at the faint end of the LF, both succeeded to mimic the steeper LF
for fainter galaxies of the observational data.  Finally,
SAG$^{\rm  b}$ overpredicts the bright end of the LF, but matches well the
observed LF for galaxies fainter than $M^{\ast}$. 
It is worth mentioning that it has been tested that adopting luminosity limited samples instead of stellar mass limited samples does not change the conclusions reached in this work.
The use of flux limited catalogues presented in the next section makes that the lower end of the SMF and the fainter end of the LF have little impact on the results.

\begin{figure}
\begin{center}
\includegraphics[width=\hsize]{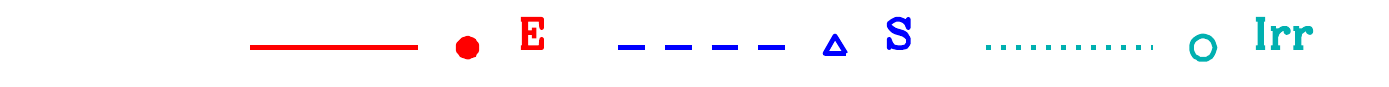}
\includegraphics[width=\hsize,viewport=0 30 400 550]{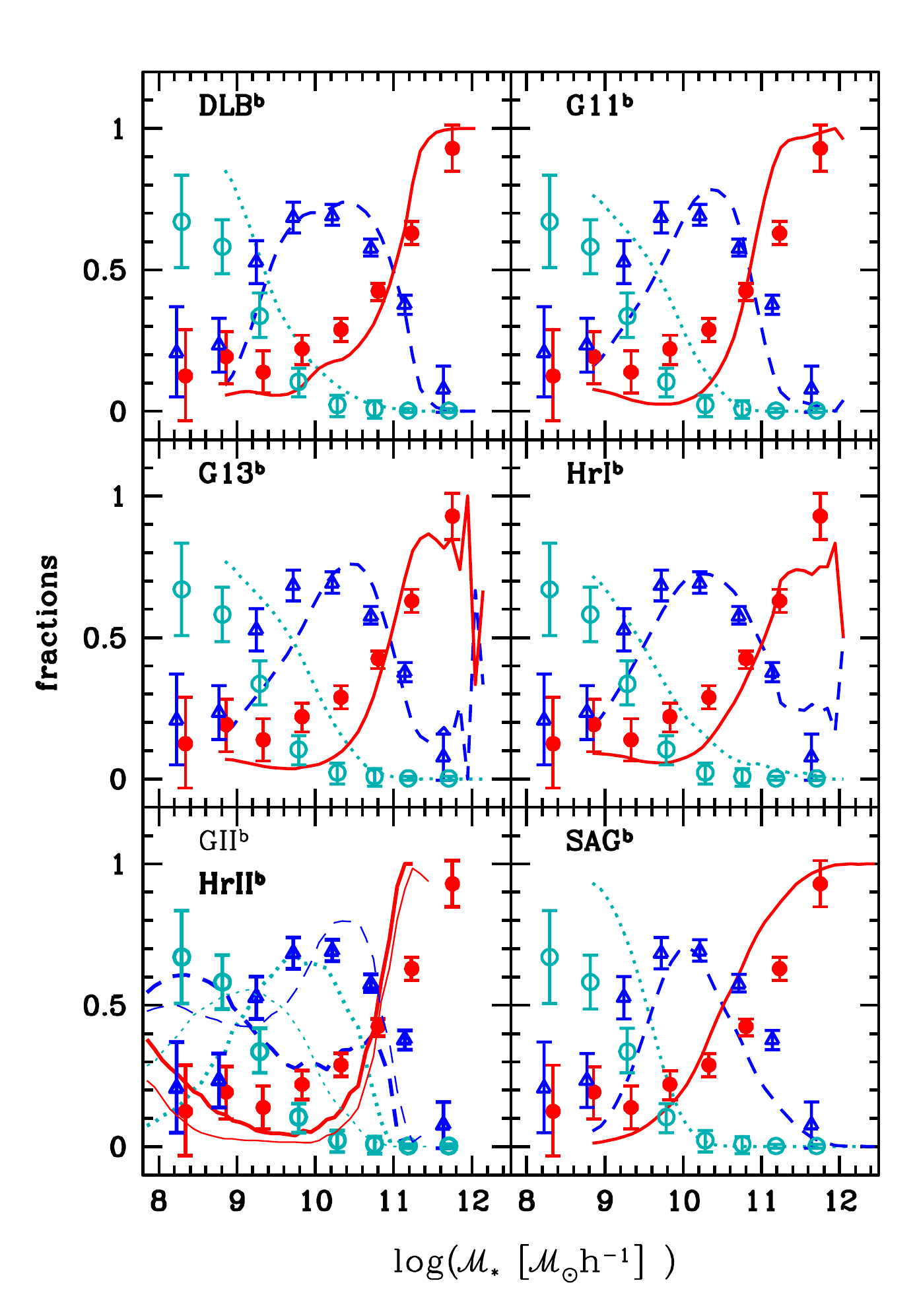}
\caption{Fractional contribution of morphological types (elliptical
  (\emph{red}), spiral (\emph{blue}), and irregular
  (\emph{cyan}), as a
  function of stellar mass in one simulation box at $z=0$. \emph{Lines} show the
  fractions obtained for the different semi-analytical models of galaxy
  formation (labelled according to acronyms in Table~\ref{tab:sams}) for the
  ellipticals (\emph{solid}), spirals (\emph{dashed}) and irregulars
  (\emph{dotted}).  
  The observations (\emph{symbols}) were computed by
  Conselice et al. (2006)
  from 
  the RC3.
}
\label{fig:fdem}
\end{center}
\end{figure}
\nocite{Conselice+06}

Figure~\ref{fig:fdem} shows the morphological fraction of
galaxies as a function of stellar masses in each SAM compared to those
obtained by \cite{Conselice+06} for galaxies in the Third Reference Catalogue
of Bright Galaxies (RC3, \citealt{RC3a}). Morphological types for synthetic
galaxies have been defined according to the bulge to total stellar mass
ratio, as follows:
 \begin{description}
\item[Elliptical galaxies:] ${\cal M}_{\rm bulge}/{\cal M}^* \ge E_{\rm lim}$, 
\item[Spiral galaxies:] $S_{\rm lim} < {\cal M}_{\rm bulge}/{\cal M}^* < E_{\rm lim}$, and
\item[Irregular galaxies:] ${\cal M}_{\rm bulge}/{\cal M}^* \leq S_{\rm lim}$,
\end{description}
We adopted limiting values $E_{\rm lim}$ and $S_{\rm lim}$ to obtain the best
recovery of the observational data: $S_{\rm lim}=0.03$ and $E_{\rm lim}=0.7$
for DLB$^{\rm b}$, G11$^{\rm b}$, G13$^{\rm b}$, GII$^{\rm b}$ and HrII$^{\rm
  b}$s \citep{Guo+11}, $S_{\rm lim}=0$ and $E_{\rm lim}=0.7$ for HrI$^{\rm
  b}$ \citep{Bertone+07}, while $S_{\rm lim}=0$ and $E_{\rm lim}=0.85 $ for
SAG$^{\rm b}$ \citep{Cora+18}. 
The MS SAMs reproduce fairly similar morphological mixes as observed.
On the other hand, GII$^{\rm b}$ and HrII$^{\rm b}$ tend to overpredict the fraction of spiral galaxies for low stellar masses (from $8$ to $9$ \Massunh), HrII$^{\rm b}$ underpredicts the fraction of spiral galaxies at intermediate masses, while SAG$^{\rm b}$ underpredicts the fraction of spiral galaxies at the highest and lowest ranges of stellar masses.

\begin{figure}
\begin{center}
\includegraphics[width=8cm,viewport=0 30 400 560]{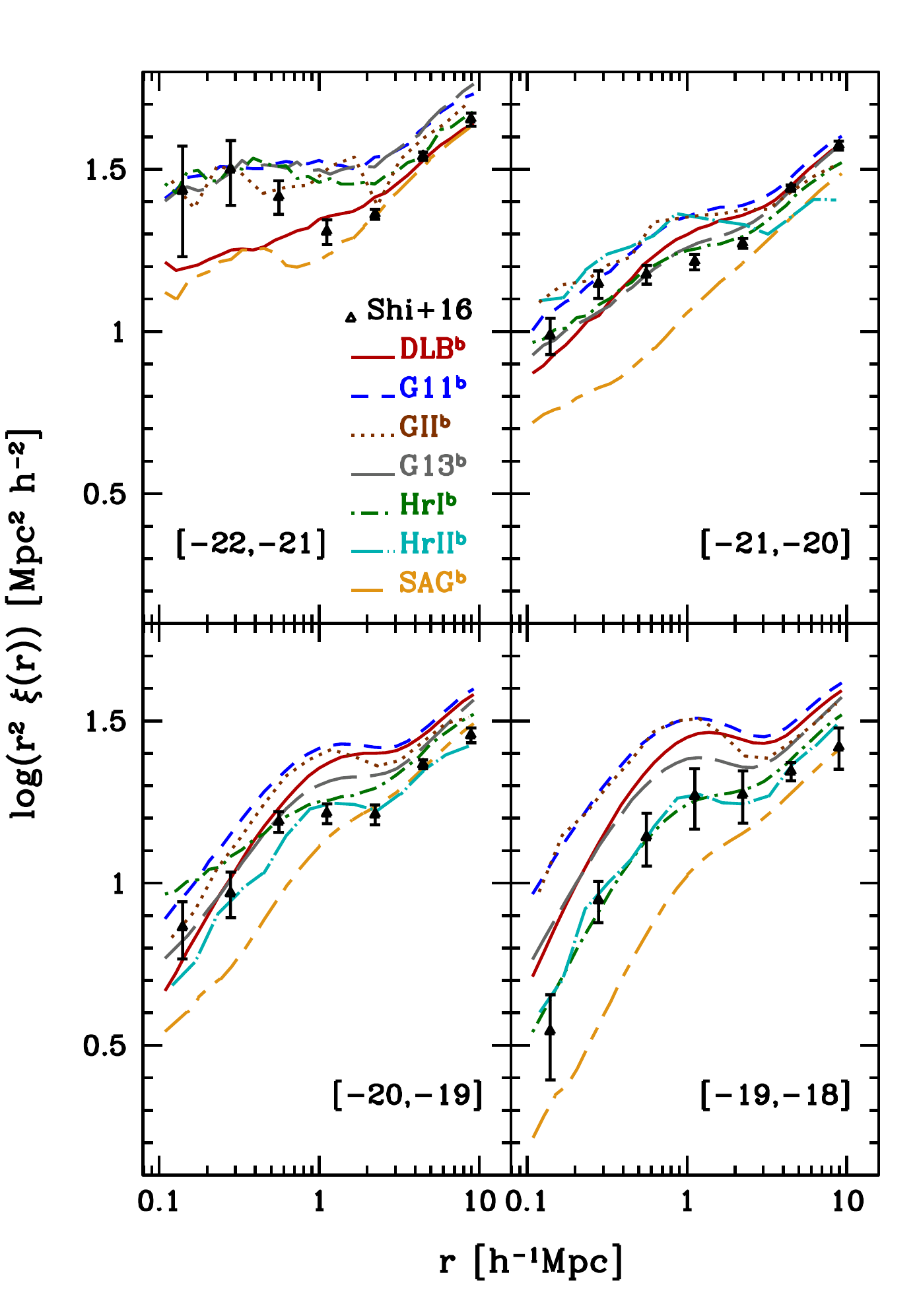}
\caption{Real-space galaxy two-point correlation functions in the $z$=0
  simulation box for the different SAMs (\emph{lines}).
Each SAM sample was restricted to match
the limits of the observational sample shown between brackets in this plot
(in the top left panel the sample HrII$^{\rm b}$ was not included given the
small number of galaxies in that magnitude range). 
The \emph{symbols} correspond to the re-constructed real-space correlation function 
for the SDSS DR7 galaxies determined by
Shi et al. (2016).
}
\label{fig:xi}
\end{center}
\end{figure}
\nocite{shi16}

In our final illustration of SAMs, Figure~\ref{fig:xi} shows the
two-point correlation functions of galaxies at $z=0$ predicted by the SAMs
computed with the estimator developed by \cite{Landy&Szalay93}
using the \texttt{tpcf} function in the package \texttt{Halotools}
\citep{halotools17}. 
We split the sample of galaxies in bins of absolute
magnitude to compare with the re-constructed real-space two-point
correlation functions computed by \cite{shi16} from SDSS DR7 galaxies
\citep{sdssdr7}.  This comparison shows that the SAMs of
\cite{Henriques+15} (HrI$^{\rm b}$ and HrII$^{\rm b}$) provide the best
recovery of the observed correlation functions in most of the magnitude ranges for
the full range of scales.  
In contrast, SAG$^{\rm b}$ underpredicts the clustering of
galaxies at small scales ($ r < 1 \, h^{-1} \, \rm Mpc$) in all the magnitude
ranges, while G11$^{\rm b}$ overpredicts the clustering in those scales and
beyond.  G13$^{\rm b}$ and DLB$^{\rm b}$ perform rather well at different
scales depending on the magnitude range.  Note that none of the
SAMs used in this work have used the correlation function to calibrate their
results. 

Differences in the recipes to build halo merger histories,  and in the treatment of the orphan galaxies (hence the
fraction of these galaxies in each $z$=0 box of the SAM) has a direct impact in clustering,
mainly at small scales  \citep{Contreras13, pujol17}, and thus could be the
reason for the differences observed in this figure. 
We examined the fraction of orphan galaxies in the $z$=0 boxes in each of the four absolute magnitude bins shown in Fig~\ref{fig:xi}. We found that, in all four magnitude bins,  SAG$^b$ has a smaller fraction of orphan galaxies than the MS-based SAMs. In Appendix~\ref{sec:orphans}, we analyse the orphan population in each box and show, in Fig.~\ref{fig:orphan_mag},  the fraction of orphan galaxies for each SAM as a function of absolute magnitude and stellar mass.

\subsection{Friends-of-Friends groups}
We also identified normal groups of galaxies in real space in each output of
the simulation boxes by means of a friends-of-friends algorithm
\citep{davis85}, which basically links galaxies whose 3D inter-particle
distances are less than a given linking length $D(z)$.
The linking length is
related to the overdensity of virialised halos relative to the mean density
of the Universe, which is a function of the
cosmology and redshift. The enclosed overdensity of halos was taken from
\cite{weinberg+03}:

\[ \Delta_{\rm vir} (z) = 18 \pi^2 \left[ 1 + 0.399\left( \frac{1}
{\Omega_{\rm m}(z)} - 1 \right)^{0.941}\right] \]
with
$$ \left( \frac{1}{\Omega_{\rm m}(z)} - 1 \right) = \left( \frac{1}{\Omega_0}-1 \right) 
(1+z)^{-3} $$
For the different cosmologies used in this work, the enclosed overdensity of virialised halos at $z=0$ are  $377$ for WMAP1, $358$ for WMAP7 and $328$ for Planck. The relation between the contour overdensity and the enclosed overdensity of virialised halos is defined by \cite{Courtin+11} as:  
\[ 
\label{rho}
\frac{\delta \rho}{\rho}(z) = b^{-3}(z) = b^{-3}_{0} 
\left(0.24 \frac{\Delta_{\rm vir}(z)}{178} + 0.68  \right) 
\] 
Following  \cite{Zandivarez+14} , we adopted  $b_0=0.14$.  
Finally, the linking length used to find neighbours is computed as 
$\displaystyle D(z)=b(z)\, l_{\rm box}/N_{\rm gal}^{1/3}$, where $l_{\rm box}$ is the box size and $N_{\rm gal}$ is the number of galaxies in the box. 
For instance, the values of the linking lengths at $z=0$ are $D(0)= \ 322,
\ 363, \ 251, \ 372, \ 401, \ 278, \ {\rm and} \ 394 \  h^{-1} \, \rm kpc$ for DLB$^{\rm b}$,
G11$^{\rm b}$, GII$^{\rm b}$, G13$^{\rm b}$, HrI$^{\rm b}$, HrII$^{\rm b}$ and SAG$^{\rm b}$,
respectively.   
The galaxy groups identified in the boxes will be used in this work only as proxies to classify different families of CGs. 

\section{Mock galaxy lightcones}
\label{sec:lightcones}

For each of the galaxy catalogues listed in Table~\ref{tab:sams}, we
constructed all-sky (i.e. $4 \, \pi\,\rm sr$) mock lightcones, following a similar
procedure as \cite{jpas}. 
The lightcones were built from the synthetic galaxies extracted at different redshifts from different
outputs of the simulations to include the evolution of structures and galaxy properties with time, and having into account for repeated or missing galaxies caused by movements between different snapshots. 
Absolute magnitudes were interpolated between different snapshots. 
Since the SAMs provide the galaxy absolute
magnitudes, to compute the observer-frame apparent magnitudes it is necessary
to include a k-decorrection procedure, which was done following the recipes
described in \cite{DiazGimenez+18}.

\begin{figure}
\begin{center}
\includegraphics[width=8cm]{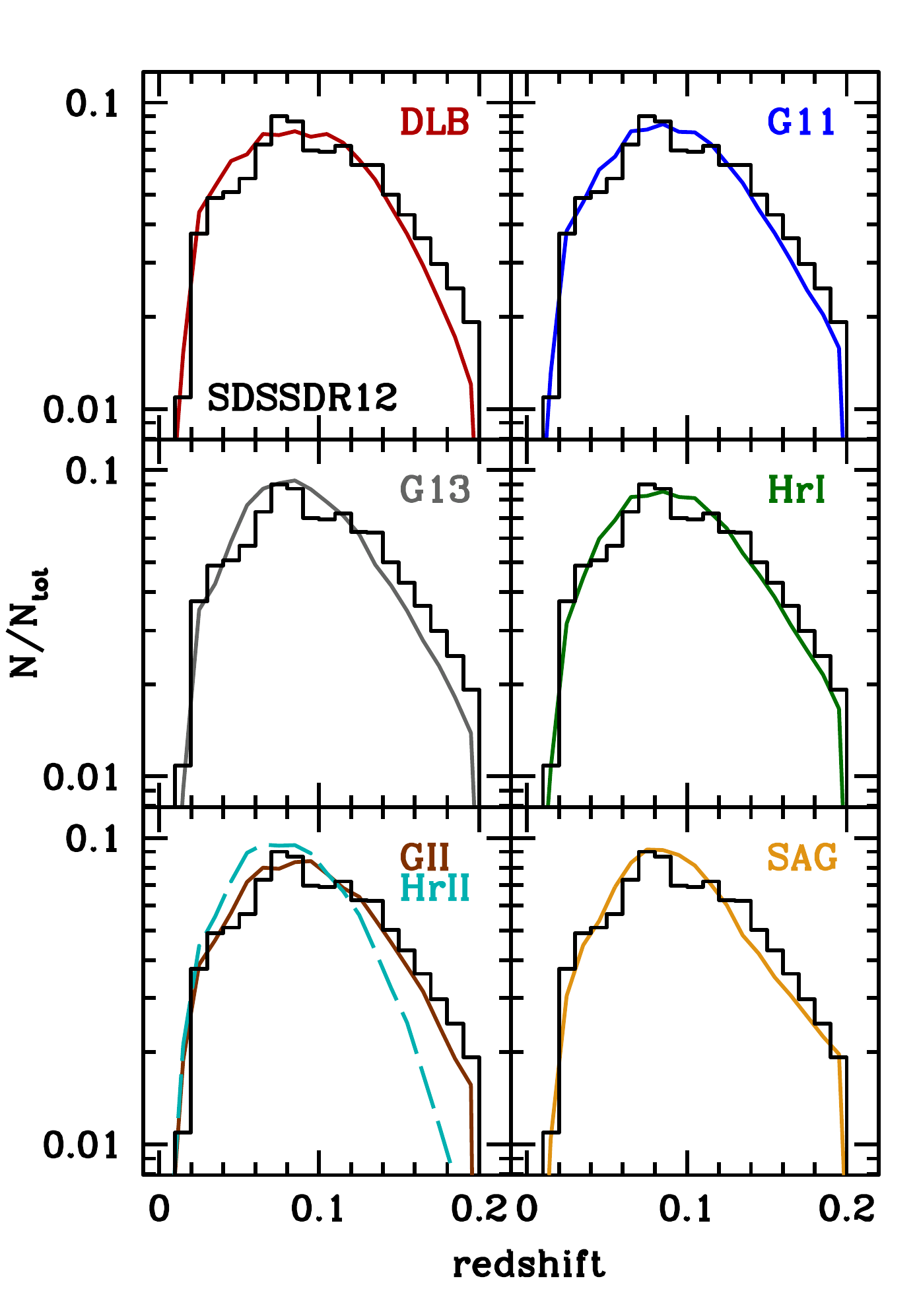}
\caption{ 
Normalized redshift distributions of galaxies in the lightcones limited to
observer frame apparent magnitude $r_{\lim}=17.77$. For the sake of comparison, lightcones (\emph{curves}) were also limited to redshift of $z_{\lim}=0.2$. \emph{Black histograms} corresponds to 
 the SDSS DR12 data compiled by
Tempel et al. (2017) using the same magnitude and redshift limits .
}
\label{fig:zdist}
\end{center}
\end{figure}
\nocite{tempel17}

\begin{table}
\begin{center} 
\caption{CGs identified in lightcones 
\label{tab:cgs}
}
\tabcolsep 2pt
\begin{tabular}{lcccccc}
\hline
\hline
Sample & \multicolumn{3}{c}{Lightcones} & &  \\
\cline{2-4}
 & photo & mag. & \# of & &\# of \\
 &  bands & limit & galaxies & & CGs \\
\hline
DLB  & $r$ & 17.77 & 3\,757\,143    && 3387   \\
G11  & $r$ & 17.77 & 3\,149\,024    && 3175   \\
GII  & $r$ & 17.77 & 3\,214\,602    && 2558  \\
G13  & $r$ & 17.77 & 2\,982\,462    && 1682   \\
HrI  & $r$ & 17.77 & 3\,087\,401    && 1291   \\
HrII & $r$ & 17.77 & 2\,519\,119    &&\ \,800 \\
HkI  & $K_{\rm s}$ & 13.57 & \ \,632\,224 && \ \,251\\
HrIc & $r$ & 16.54 & \ \,611\,008 && \ \,276\\
SAG  & $r$ & 17.77 & 2\,941\,613    &&\ \,723 \\
\hline
cDLB	& $r$ & 17.77 & 3\,757\,143   &&1812    \\
cHrI 	& $r$ & 17.77 & 3\,087\,401   &&\ \,684 \\
\hline
SDSS & $r$ & 17.77 & \ \,557\,517 &  & \ \,462 \\
c2MASS & $K_{\rm s}$ & 13.57 & \ \,408\,618 &  & \ \ \,85 \\
\hline
\end{tabular}  
\end{center} 
\parbox{8cm}{
HkI: the same as HrI but built using the $K_{\rm s}$ photometric band.
HrIc is built similarly to HrI but with a different apparent magnitude limit.
The numbers of CGs are deduced from the new CG finder of \cite{DiazGimenez+18}.
cDLB: \texttt{classic} CG finder on DLB lightcone;
cHrI: \texttt{classic} CG finder on HrI lightcone.
SDSS: observational results from \cite{DiazGimenez+18} within a solid angle of $0.66 \, \pi\,\rm sr$;
c2MASS: observational results from \cite{DiazGimenez+12} within a solid angle of $2.3 \, \pi\,\rm sr$. 
}
\end{table}
\begin{figure*}
\centering
\includegraphics[width=0.45\hsize,viewport=15 20 330 504]{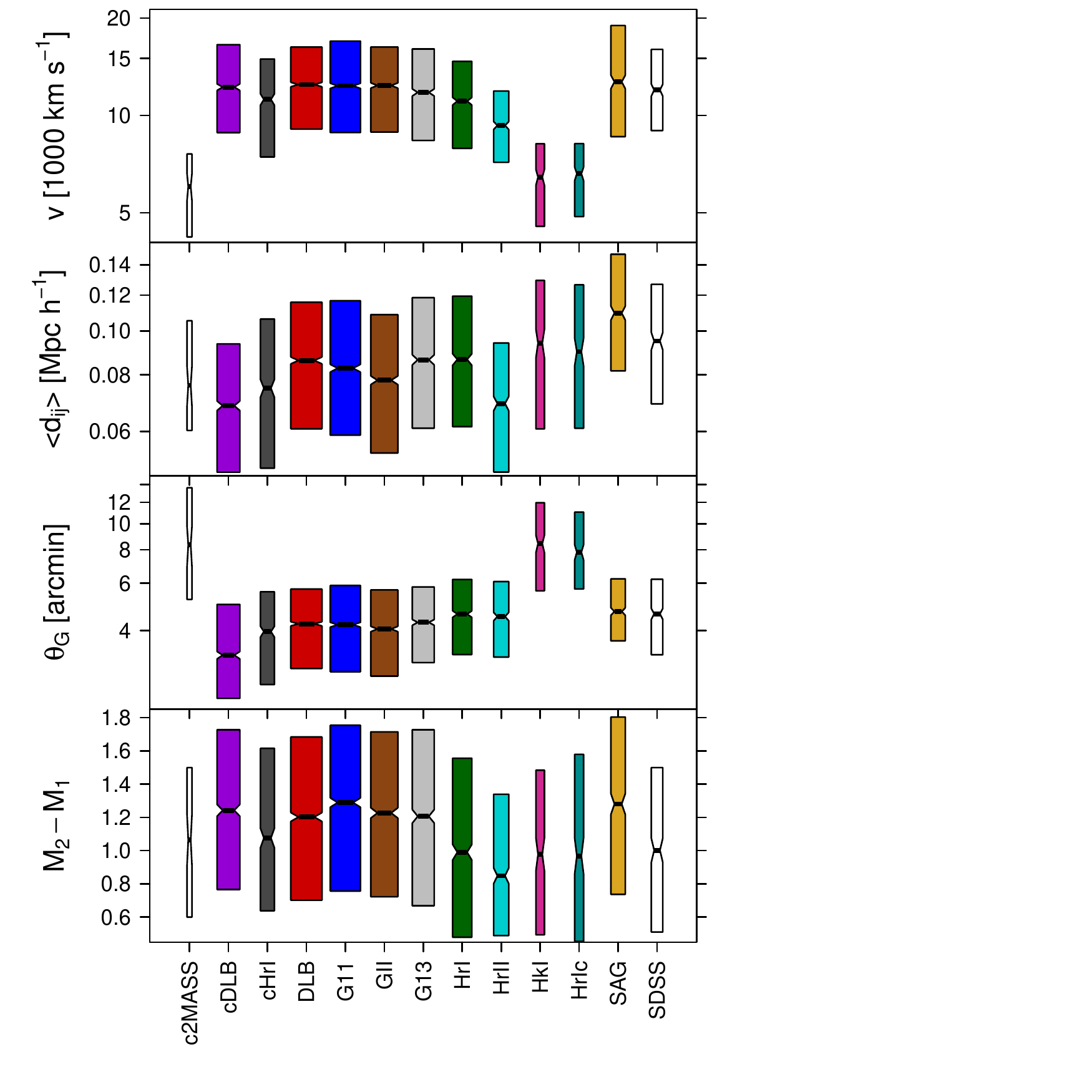}
\includegraphics[width=0.45\hsize,viewport=15 20 330 504]{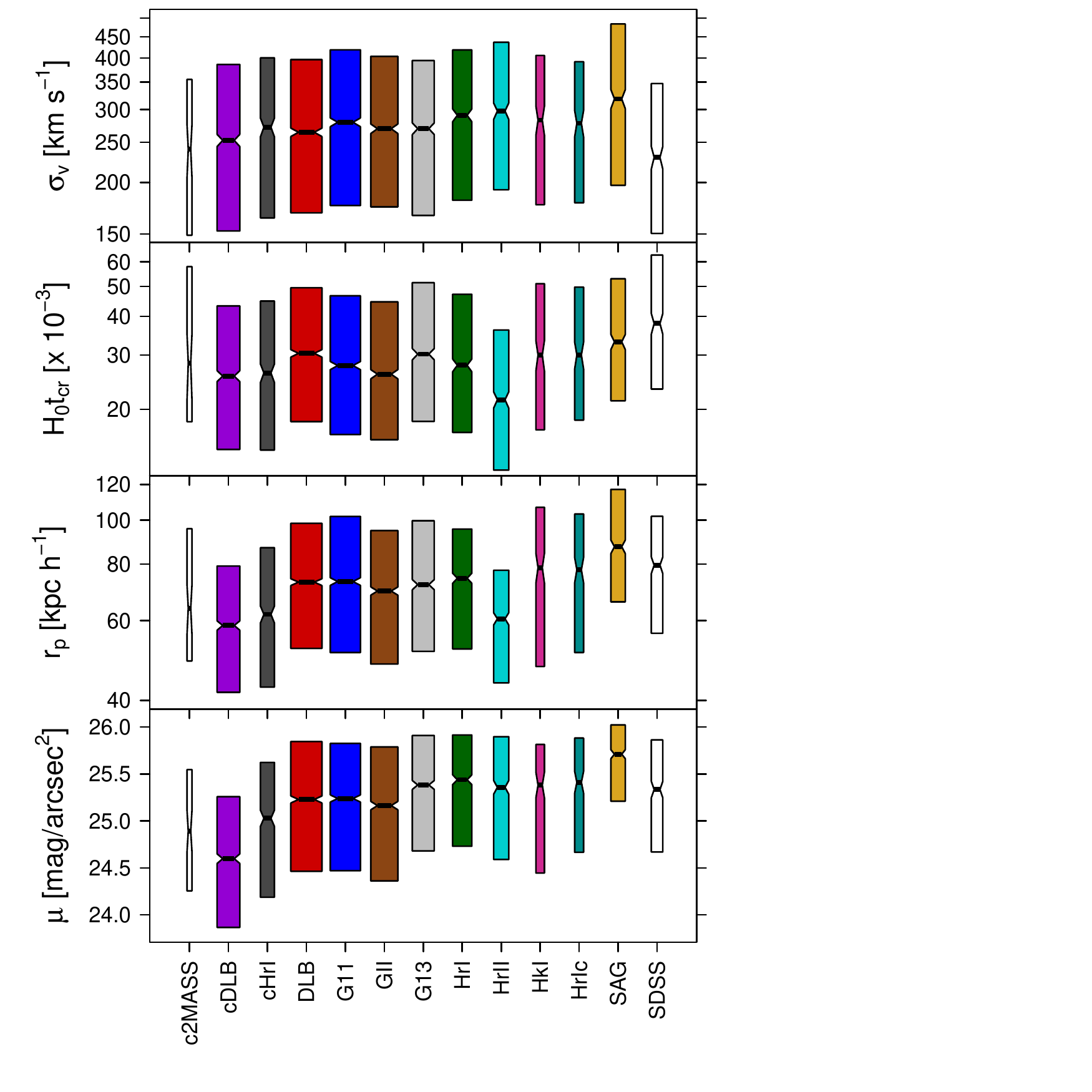}
\caption{\label{fig:distr} Distributions of properties of CGs identified in
  different lightcones.
  \emph{Left, top to bottom}: heliocentric velocity, median projected intergalactic
  separation, angular diameter, magnitude gap between the two brightest
  galaxies.
  \emph{Right, top to bottom}: line-of-sight velocity dispersion estimated via the gapper estimator (Beers et al. 1990), 
  crossing time, projected radius, mean surface brightness
  (for samples identified in the $K_{\rm s}$ band, the surface brightness has
  been increased by $2.73$ magnitudes to compare the $r$ band of the other SAMs). 
  The waists, tops and bottoms of the coloured boxes
  indicate the median values, and the 75th and 25th percentiles,
  respectively, the widths of the boxes are proportional to the numbers of CGs, while notches show the $95\%$ confidence interval.
  For comparison, we also show the CGs properties for two
  observational samples: the SDSS (identified by
D\'{\i}az-Gim\'enez et al. (2018)  
  with the new CG finder) and the 2MASS (identified by
D\'{\i}az-Gim\'enez et al. (2012)  
with the classic CG finder). In Table~\ref{tab:pvalues}, we show how  SAMs compare to observational catalogues. 
}
\end{figure*}
\nocite{DiazGimenez+12}
\nocite{DiazGimenez+18}
\nocite{Beers+90}

We built nine
different lightcones using different parent dark matter - only simulations,
resolutions, SAMs, wavebands and depths to understand the
influence of the cosmological parameters as well as resolution, CG algorithm etc..
Most of our lightcones are limited to an apparent
observer-frame SDSS AB magnitude of $r \leq 17.77$. But we also built
a lightcone by selecting galaxies in the $K_{\rm s}$ band\footnote{we
  converted the magnitudes in the SAM from the AB to Vega system to mimic the
  2MASS magnitudes} with apparent magnitude limit of $K_{\rm s} = 13.57$. Another
 lightcone is limited to $r\le 16.54$, which is the equivalent limit
in the $r$ band to the Two Micron All Sky Survey (2MASS) $K_{\rm s}$
limit. 
The total number of galaxies in each lightcone
is quoted in Table~\ref{tab:cgs}. Figure~\ref{fig:zdist} shows a comparison
of the normalised redshift distribution of each of the lightcones with
apparent magnitude limit of $r < 17.77$ and redshift\footnote{the redshift limit was imposed only to build this figure to perform a fair comparison with the observational data} $z \leq 0.2$ with the observational redshift
distribution of the galaxies in the SDSS DR12 compiled by \cite{tempel17}.
The 7 lightcones built from different SAMs all lead to a good match of the
redshift distribution, except that they all underestimate the numbers of
galaxies in the range $0.14 < z < 0.2$. 

\section{The compact group samples}
\label{sec:sample}

We identified mock CGs in the galaxy lightcones following the basic criteria
defined by \cite{Hickson82,Hickson92}. We used the new CG finding algorithm
of \cite{DiazGimenez+18}.  Basically, the CG finder
looks for groups that obey the following constraints:
\begin{description}
    \item \noindent {\bf [membership]} between 4 to 10 members within a magnitude range of three from their brightest galaxy ($4 \le N(\Delta m <3) \leq 10$); 
    \item \noindent {\bf [flux limit]} group brightest galaxy  at least 3 magnitudes
      brighter than the limiting magnitude of the parent catalogue ($m_{\rm
        b} \le m_{\rm lim} - 3$) to ensure that all CGs span the $m_{\rm
        f}-m_{\rm b} = 3$ range of magnitudes.
    \item \noindent {\bf [compactness]} mean surface brightness averaged over the smallest circle that
      circumscribes the member galaxy centres above threshold $\mu_G \le
      \mu_{\rm lim}$; 
    \item \noindent {\bf [isolation]} no galaxies in the
      same range of redshift of the galaxy members, within the considered
      magnitude range or brighter, and within three times the size of the
      smallest circumscribed circle ($\Theta_n > 3 \,\Theta_G$);
    \item \noindent {\bf [velocity concordance]} all the galaxy members with concordant radial velocities (less than
      $1000 \, \rm km \, s^{-1}$ from the median radial velocity of the group centre); 
\end{description}

The value of $\mu_{\rm lim}$ (surface brightness limit) depends on the photometric band in which the selection is made. We worked in two different wavebands, and therefore these values vary accordingly. For catalogues in the SDSS-r band, we adopted $\mu_{\rm lim} =26.33$, while the equivalent value for the sample identified in the $K_{\rm s}$ band is $\mu_{\rm lim} =23.6$ (for a full description, see \citealt{Taverna+16}). The limiting magnitudes, $m_{\rm lim}$, are quoted in Table~\ref{tab:cgs}.

Until recently, Hickson-like CGs have been usually
identified in two steps: first, groups were identified in projection in the
plane of sky; and second, a velocity filter was applied to the projected
groups. This method has the drawback that CGs with distant background or
foreground  galaxies lying just
outside are rejected, leading to highly incomplete samples.
Our new CG finder (\citealp{DiazGimenez+18}, DGZT18) identifies CGs directly in
2+1D redshift space, thus avoiding losing such CGs, which more than doubles
the completeness (see DGZT18).
In the present study, we focus on the new CG finder, but also show results
using the so-called {\tt classic} algorithm, 
using a `c' prefix to name the samples obtained with that algorithm.

While galaxies in the mock lightcones are just point-sized particles,
observed galaxies are extended objects. Following DGM10, we have therefore
included the blending of galaxies in projection on the plane of the sky,
which modifies the number of detectable galaxies and changes the population
of CGs. We computed their half-light radii in the $r$ and $K_{\rm s}$ band as
a function of the stellar mass of each mock galaxy following the
prescriptions of \cite{Lange+15} for different morphological types.  Finally,
we considered that two galaxies are blended if their angular separation is
smaller than the sum of their angular half-light radii (DGM10). The number of
CGs identified in each lightcone is quoted in Table~\ref{tab:cgs}.  

Figure~\ref{fig:distr} shows the distribution of different properties of CGs
in all the lightcones in form of boxplot diagrams
\citep{boxplot78,boxplot14}.
We also included the properties for CGs extracted in the $K_{\rm s}$ band from the 2MASS catalogue by \cite{DiazGimenez+12} using the classic algorithm (labelled as c2MASS), and the sample of CGs identified by \cite{DiazGimenez+18} using the newest version of the finder algorithm in the $r$ band in the SDSS DR12 (labelled as SDSS).
Two of the properties shown in this figure are explicitly dependent on the distance ($v$, $\Theta_{\rm G})$, therefore, the samples extracted from HkI, HrIc and c2MASS show differences in these properties by construction. 

\begin{figure*}
\begin{center}
\includegraphics[width=0.8\hsize,viewport=10 25 554 554]{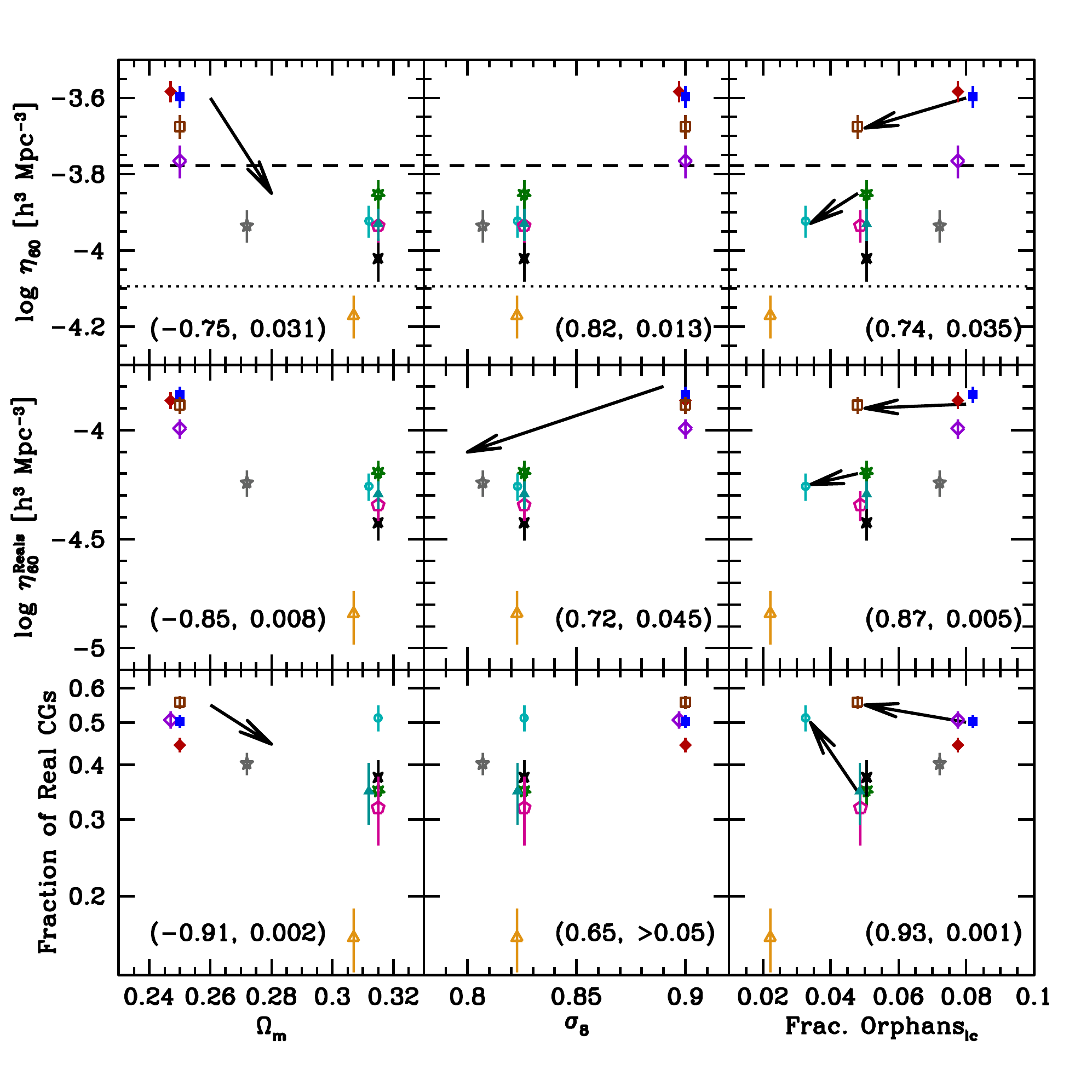}
\includegraphics[width=0.9\hsize,viewport=10 0 480 60]{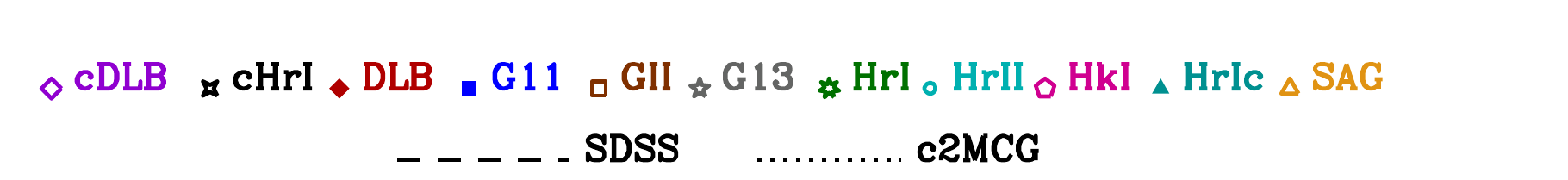}
\caption{\label{fig:fcosmo}
Space densities of CGs (\emph{upper panels}) and of Real CGs
  (\emph{middle panels}), and fraction of Real CGs (\emph{lower panels}) as a
  function of the cosmological matter density parameter
  ($\Omega_{\rm m}$) (\emph{left panels}), the amplitude of the mass fluctuations in a sphere
  of $8 \,h^{-1}\, \rm Mpc$ ($\sigma_8$) (\emph{middle panels}), and the fraction of orphan galaxies in the lightcones (\emph{right panels}). 
  Each SAM is labelled with their
  corresponding acronyms defined in Table~\ref{tab:cgs}.  In the \emph{upper
    panels}, the \emph{horizontal dashed lines} correspond to the space
  density of the observational sample obtained from SDSS DR12 by
  \citealt{DiazGimenez+18}, while the dotted lines correspond to the sample
  identified with the classic finder from 2MASSXCG by
  \citealt{DiazGimenez+12}.
 The arrows in the left and right panels indicate the transition in cosmological parameters keeping most other things the same (i.e. from G11 to G13). The arrows in the right panels indicate the transition in resolution (G11 to GII and HrI to HrII).
 In each panel, numbers indicate $(r_{\rm S},p)$, i.e. the Spearman rank correlation coefficient ($r_{\rm S}$) and the corresponding $p$-value, computed only from the lightcones derived from SAMs built on the Millennium~I simulation. 
 }
\end{center}
\end{figure*}

For all the other properties with no dependence on the distance, from this figure, we observe 
that there is not a large spread between the physical properties computed from different SAMs and observations. 
A detailed statistical analysis (using the Kolmogorov-Smirnov two-sample test and the  95\% confidence intervals for the medians) allow us to perform a more quantitative comparison between in CG properties between SAM and observed samples (see Table~\ref{tab:pvalues}). 
From this analysis we can highlight the following results: when using the classic algorithm, both SAMs (DLB and HrI) reproduce the observational CG properties quite well; when using the modified algorithm HrI, HkI and HrIc CG samples reproduce fairly well several observational properties ($d_{ij}$, $M_2-M_1$, $r_p$ and $\mu$), but not others ($\sigma_{\rm v}$, $H_0\,t_{\rm cr}$, and $\mu$).
Also, some different behaviours can be observed when comparing SAMs vs SAMs. For instance, samples identified with the classic algorithm show lower surface brightnesses, projected radii and crossing times as a consequence of the incompleteness of the identification as has been previously discussed in \cite{DiazGimenez+18}.
CGs extracted from the \cite{Henriques+15} SAM (cHrI, HrI, HrII, HkI and HrIc) present the lowest median of the magnitude difference between the two brightest galaxies (i.e., avoid CGs dominated by a single galaxy), while SAG and G11 CGs show the largest magnitude gaps. 
Also, HrII shows the typical lowest magnitude difference, projected separation, projected radius and crossing time among all the samples, while the SAG CGs behaves the opposite in those properties.

\section{Frequency and Nature of Compact Groups}
\label{sec:results}

We measured the space density of CGs in each mock lightcone up to the median distance of the shallowest samples. We defined 
$$\eta_{60}= \frac{3 N_{\rm CG}(r<r_{60})}{\Omega \, r^3},$$ 
with a solid angle $\Omega=4\pi$  and a distance $r_{60}=60 \, h^{-1} \rm Mpc$, which is close to the median of the comoving distances of the group centres in the HkI and HrIc lightcones which have the most restrictive flux limit (see top left panel of Fig.~\ref{fig:distr}).

\begin{figure*}
\begin{center}
\includegraphics[width=0.8\hsize,viewport=10 30 495 270]{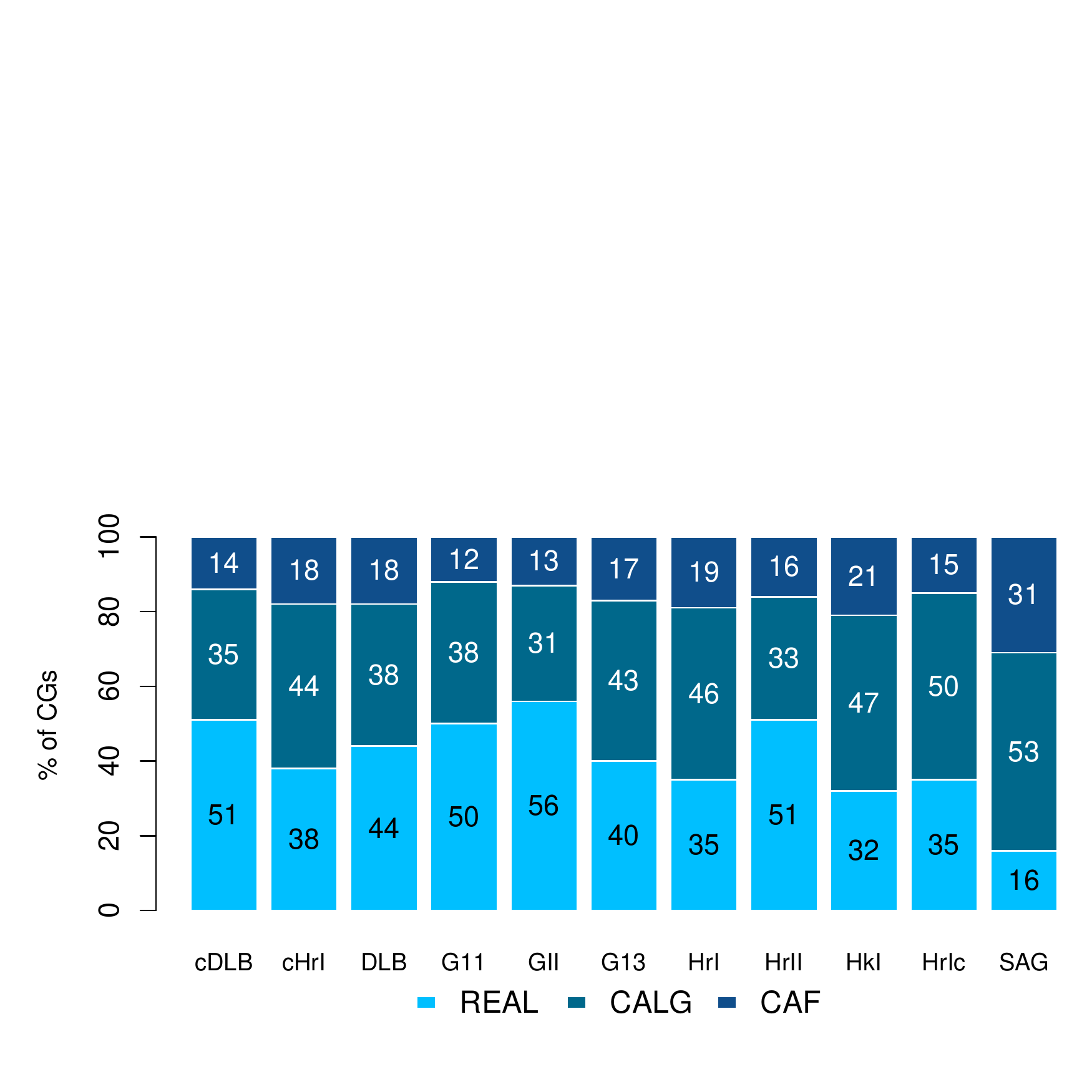} 
\includegraphics[width=0.8\hsize,viewport=10 30 495 270]{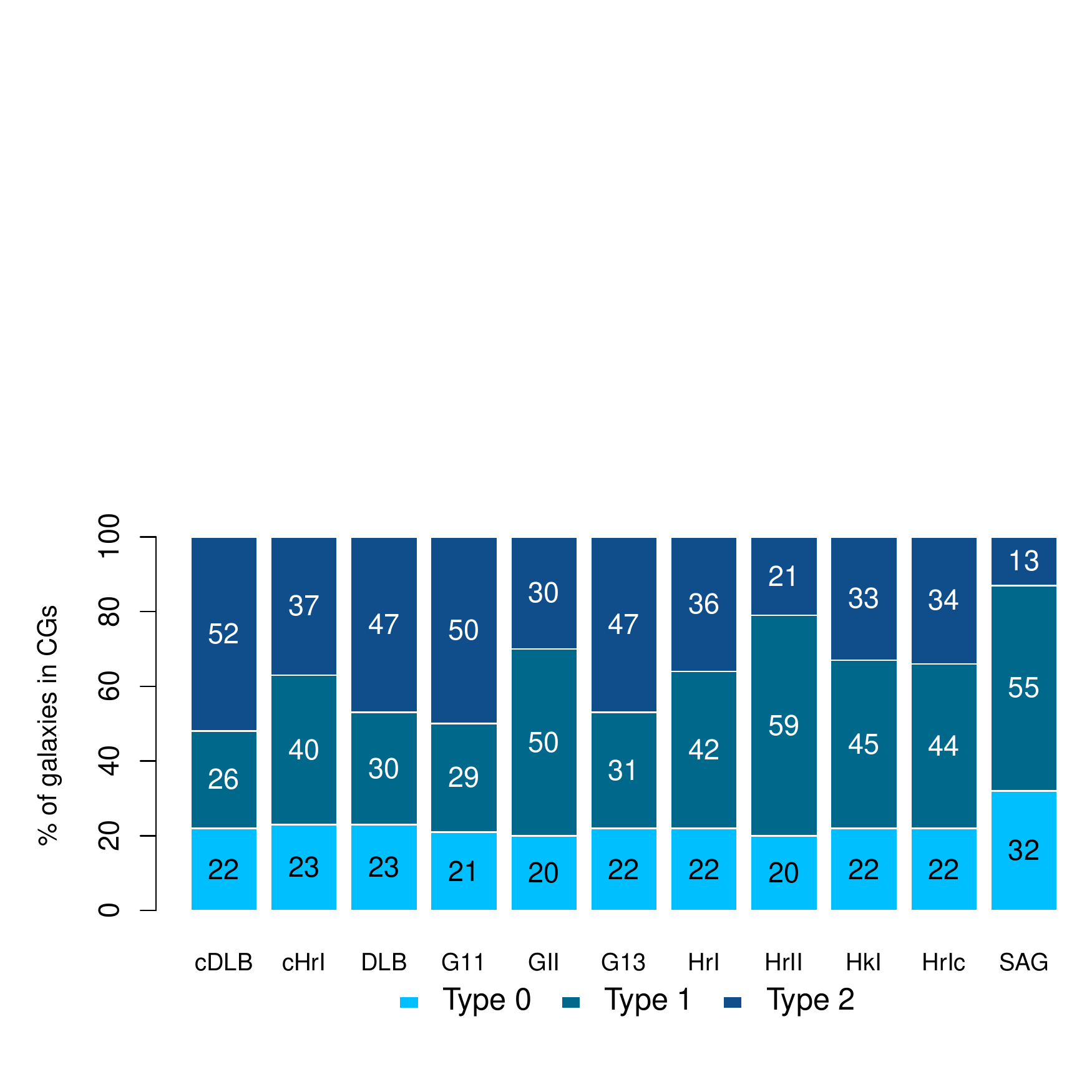}
\caption{\label{fig:stackbar} Top panel: Percentages of different classes of CGs for the
  9 SAM lightcones:
  physically dense (REAL),
  chance alignments within looser groups (CALG) and within the field (CAF).
  Bottom panel: Percentages of galaxies in CGs that are classified as central galaxy of a FoF group (type 0), central galaxy of a subhalo (type 1), or a satellite or orphan galaxy (type 2).
 }
\end{center}
\end{figure*}

Top panels of Figure~\ref{fig:fcosmo} show the space density of CGs in
each lightcone as a function of the cosmological parameters: matter density,
$\Omega_{\rm m}$ (left), and the primordial density
fluctuation amplitude at $8 \, h^{-1} \, \rm Mpc$, $\sigma_8$ (middle), and as a function of the fraction of orphan galaxies in the lightcones (right).
In the top panels, we have also included the space density of the two
observational CG samples previously described: c2MASS (dotted lines) and SDSS
(dashed lines).  Given the limitations inherent to the observational
catalogues (redshift incompleteness, fiber collisions, brightness
saturation), the completeness of those samples is compromised, therefore, the
values of space density found in observations should be taken as lower
limits.  The top panels of Fig.~\ref{fig:fcosmo} indicate
that the space density of the samples found with the classic CG finder (cDLB
and cHrI) are above the
values of the space density of the 2MASS CG sample identified with the same
algorithm. However, the values of the space densities obtained from all the
samples based on \cite{Henriques+15}, G13 and SAG using the new CG finder
are below the threshold given by SDSS CGs, therefore these SAMs
seem to be missing CGs compared to observations.

We split the samples of CGs into \emph{physically dense} (Reals) and \emph{chance
alignments} (CAs) following the classification performed by DGM10. We
classified a CG as Real if at least four of its galaxies form a physically
dense group.  To achieve this goal, DGM10 defined a classification based on
the 3D inter-particle separation and the ratio between the projected and
line-of-sight
sizes of the four closest galaxies in the CG. Therefore, we classify a
CG as Real when the 3D comoving maximum inter-particle separation between the
four closest galaxies is less than $100 \, h^{-1} \, \rm kpc$, or less than
$200 \, h^{-1} \, \rm kpc $ while the ratio of line-of-sight to transverse
sizes in real space is not higher than $2$.
Hence, all CGs that not fulfil this criterion are classified as CAs.

\subsection{Effects of the cosmological parameters}
The variation of the space density of Real CGs and the fraction of Real CGs
as a function of the cosmological parameters are shown in the middle and
bottom panels of Fig.~\ref{fig:fcosmo}. The trends of space density of the
Real CGs follow the same behaviour as the space density of the total sample
of CGs.  Considering the samples built from the same simulation (MS),
the
fraction of Reals decreases as $\Omega_{\rm m}$ increases, while there is not
obvious trend with $\sigma_8$.

The main aspects of Figure~\ref{fig:fcosmo} are seen by
comparing G11 and G13 (see arrows), which use very similar SAMs on the same Millennium
simulation, but run with WMAP7 cosmology for the latter. 
Assuming power
law trends, we find that
\begin{eqnarray}
  \eta_{60} &\propto& \Omega_{\rm m}^\alpha \ ,
  \label{etavsOmegam}\\
  \eta_{60}^{\rm Reals} &\propto& \sigma_8^\beta \ ,
  \label{etaRealvssigma8} \\
  f_{\rm Real} &\propto& \Omega_{\rm m}^\gamma \ ,
  \label{frealvsOmegam}
\end{eqnarray}
where 
\begin{eqnarray}
\alpha &=& \left \{
\begin{array}{ll}
  -9.3\pm1.4 & ({\rm G11}\to {\rm G13}) \\
  -2.7\pm0.8 & ({\rm all \ MS}) \\
  \end{array}
  \right. \ ,
  \label{alpha}\\
 \beta &=& \left\{ 
\begin{array}{ll}  
  \ \,8.5\pm1.5 & ({\rm G11}\to \rm{G13}) \\
  \ \,9.8\pm1.9 & ({\rm all \ MS})\\
  \end{array}
  \right. \ , 
  \label{beta}\\
  \gamma &=& \left\{
  \begin{array}{ll}
  -2.7\pm0.8 & ({\rm G11} \to {\rm G13}) \\
  -1.4\pm 0.2 & {\rm (all \ MS)} \\
  \end{array}
  \right. 
  \label{gamma}
  \ .
\end{eqnarray}

These statistically significant trends can be interpreted as
\vspace{-0.5\baselineskip}
\begin{enumerate}
\item an increased contamination of CG annuli with increasing $\Omega_{\rm m}$
(eqs.~[\ref{etavsOmegam}] and [\ref{alpha}]),
\item an increased space density of physically dense CGs within increasing
normalisation of the matter power spectrum (eqs.~[\ref{etaRealvssigma8}] and [\ref{beta}]),
\item an  increased fraction of CGs by chance alignments with increasing
  $\Omega_{\rm m}$ 
  (eqs.~[\ref{frealvsOmegam}] and [\ref{gamma}]).
\end{enumerate}

This assumes that the physical recipes in the G11 and G13 SAMs are
identical, which is not entirely true. Indeed,
according to table~2 of \cite{Guo+13}, relative to G11,
the star formation efficiency in G13
is half of what it is in G11,
the feedback mechanisms are half in G13 for SN mechanical
  feedback and AGN feedback, while the thermal feedback of SNe is roughly the
  same between the two SAMs.
However, our trends with increasing $\Omega_{\rm m}$ of both decreased CG space densities and decreased fractions
of physically dense CGs, as well as the increase
of the space density of physically dense CGs with increasing $\sigma_8$ are
seen for the SAMs displayed in Figure~\ref{fig:fcosmo}, and particularly, among those from the MS with the same DM resolution which are displayed in equations~(\ref{alpha}),
(\ref{beta}) and (\ref{gamma}) with `(all MS)'. 
We analysed the strength of these correlations obtained from the MS samples using the Spearman rank correlation coefficient and their $p$-value, which are quoted in Fig.~\ref{fig:fcosmo}. We found a strong anti-correlation between the total density, the density of Real CGs and the fraction of Real CGs with $\Omega_{\rm m}$, and a strong correlation between the total density and density of Real CGs with $\sigma_8$, but no significant correlation of the fraction of Real CGs with $\sigma_8$.

CAs can occur in two ways: \emph{chance alignments of galaxies within larger
(looser) groups} (CALGs) and \emph{CAs within the field}  (CAFs), i.e. filaments
viewed end-on, where the peculiar velocities roughly cancel by chance the
differences in the velocities caused by the Hubble flow to allow the CG to
meet the velocity concordance criterion.
This classification is performed by using the identification (by
\emph{Friends-of-Friends} on the 3D galaxy distribution) of normal (`FoF') groups
in the simulation boxes described at the end of Sect.~\ref{sec:simulations}: 
\vspace{-0.5\baselineskip}
\begin{itemize}
    \item if the four closest galaxies in a CA CG belong to only one FoF group, the group is classified as CALG;
    \item if the four closest CG members belong to more than one FoF group or none, then the group is classified as CAF.
\end{itemize}

\begin{figure*}
  \begin{minipage}[b]{0.44\textwidth}
    \centering
    \includegraphics[width=1.0\textwidth,trim=0 170 15 15,clip]{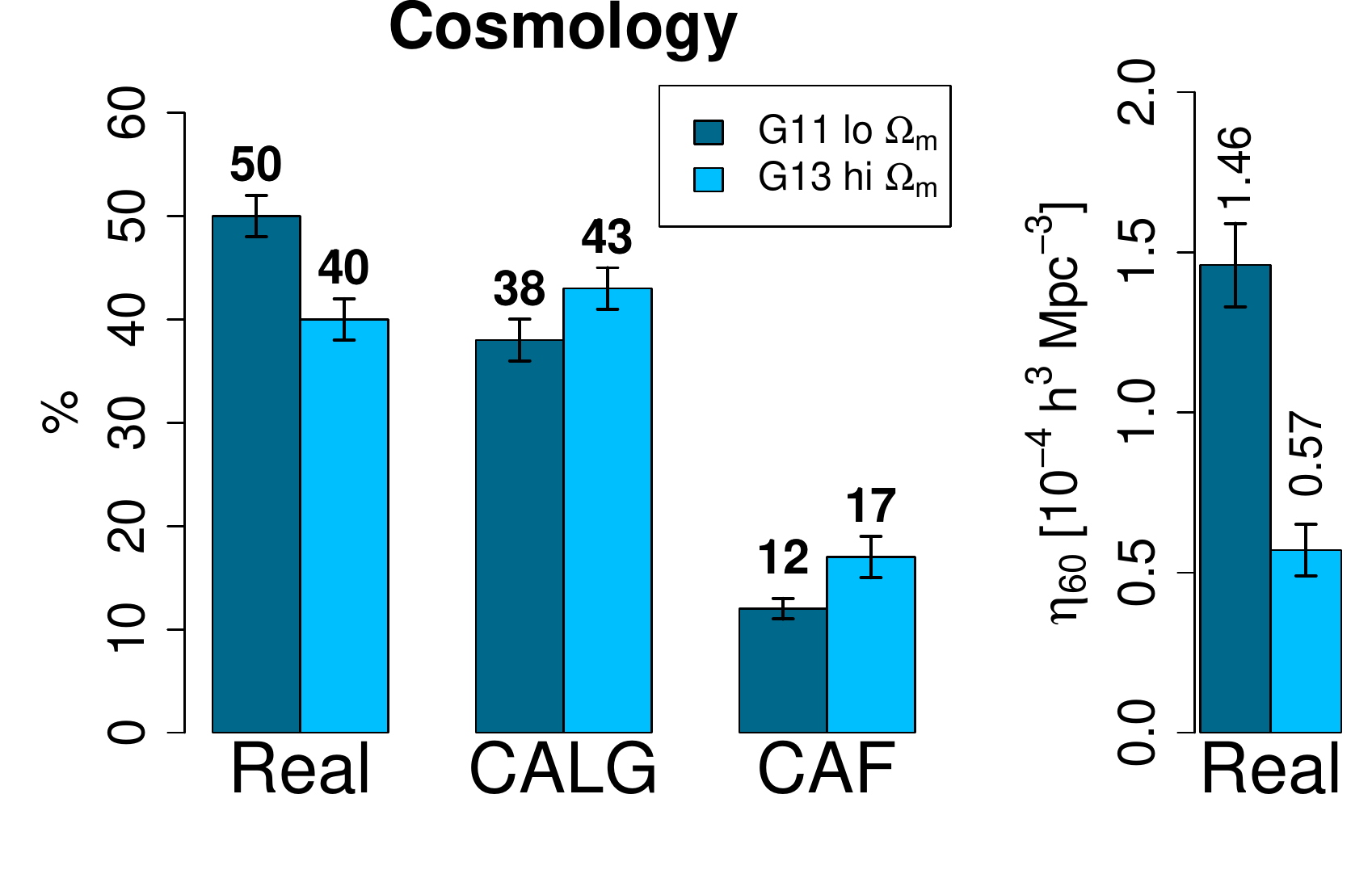}
  \end{minipage}
  \begin{minipage}[b]{0.44\textwidth}
    \centering
    \includegraphics[width=1.0\textwidth,trim=0 170 15 15,clip]{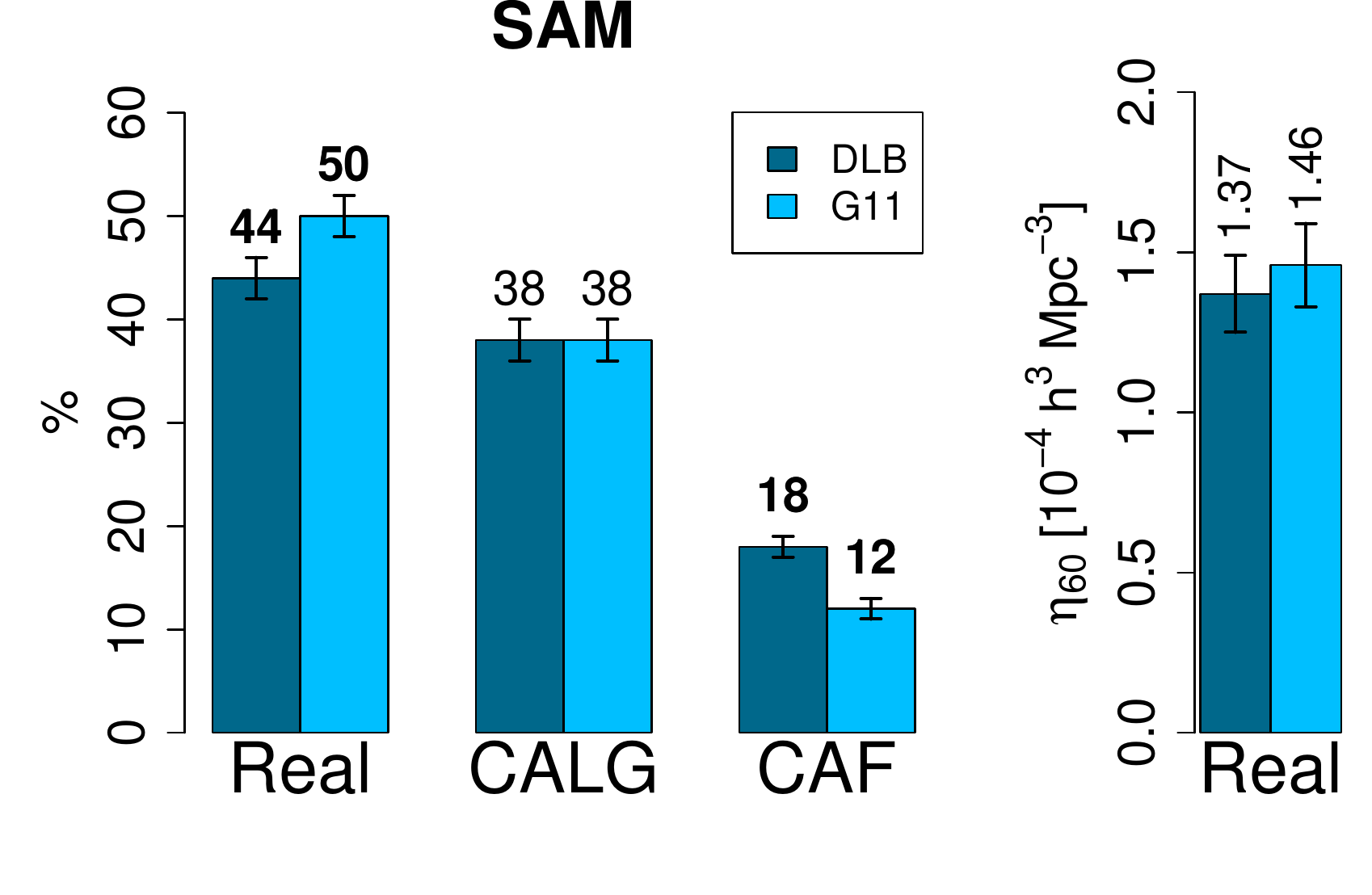}
  \end{minipage}
  \begin{minipage}[b]{0.44\textwidth}
    \centering
    \includegraphics[width=1.0\textwidth,trim=0 170 15 15,clip]{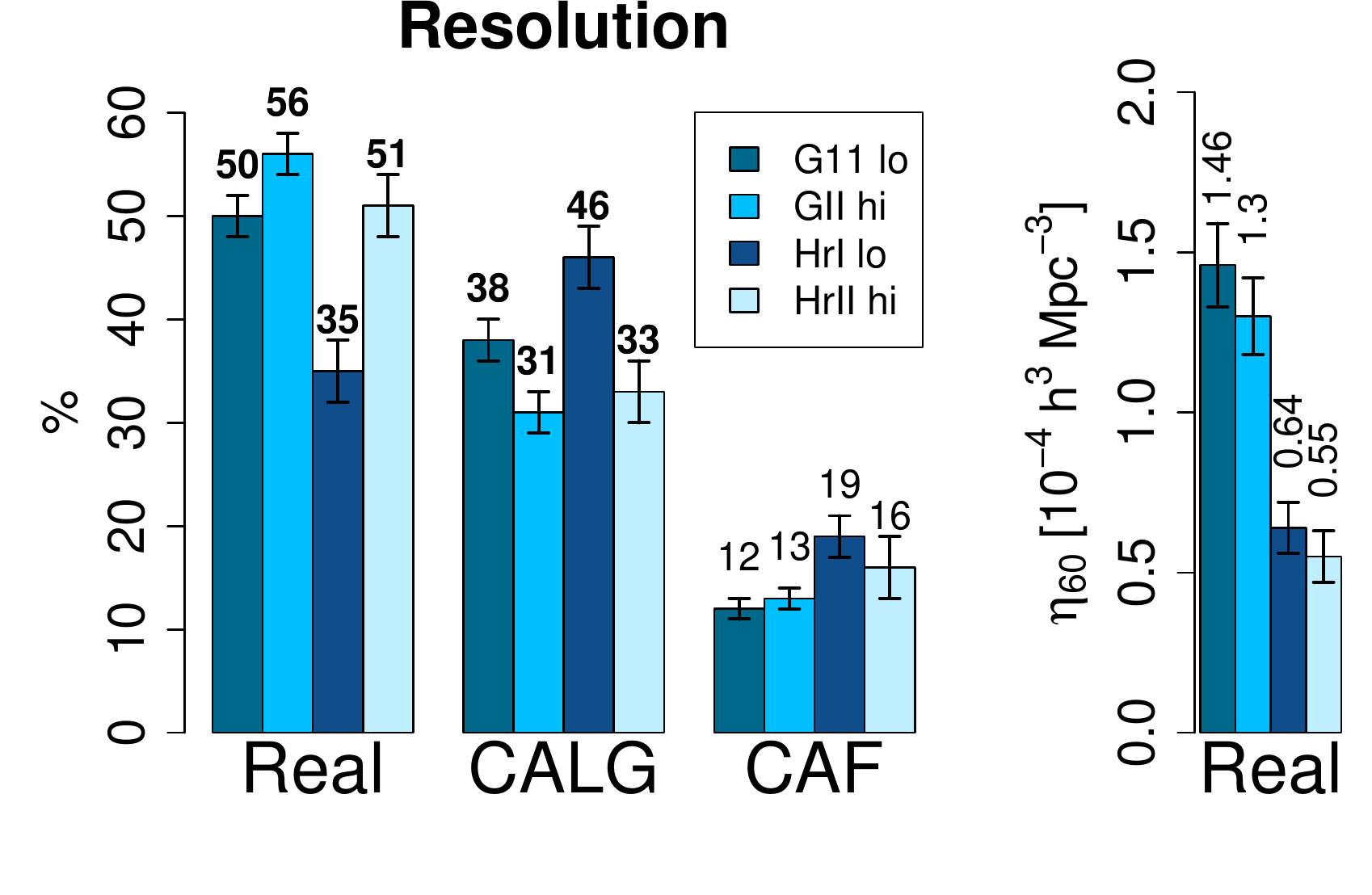}
  \end{minipage}
  \begin{minipage}[b]{0.44\textwidth}
    \centering
    \includegraphics[width=1.0\textwidth,trim=0 170 15 15,clip]{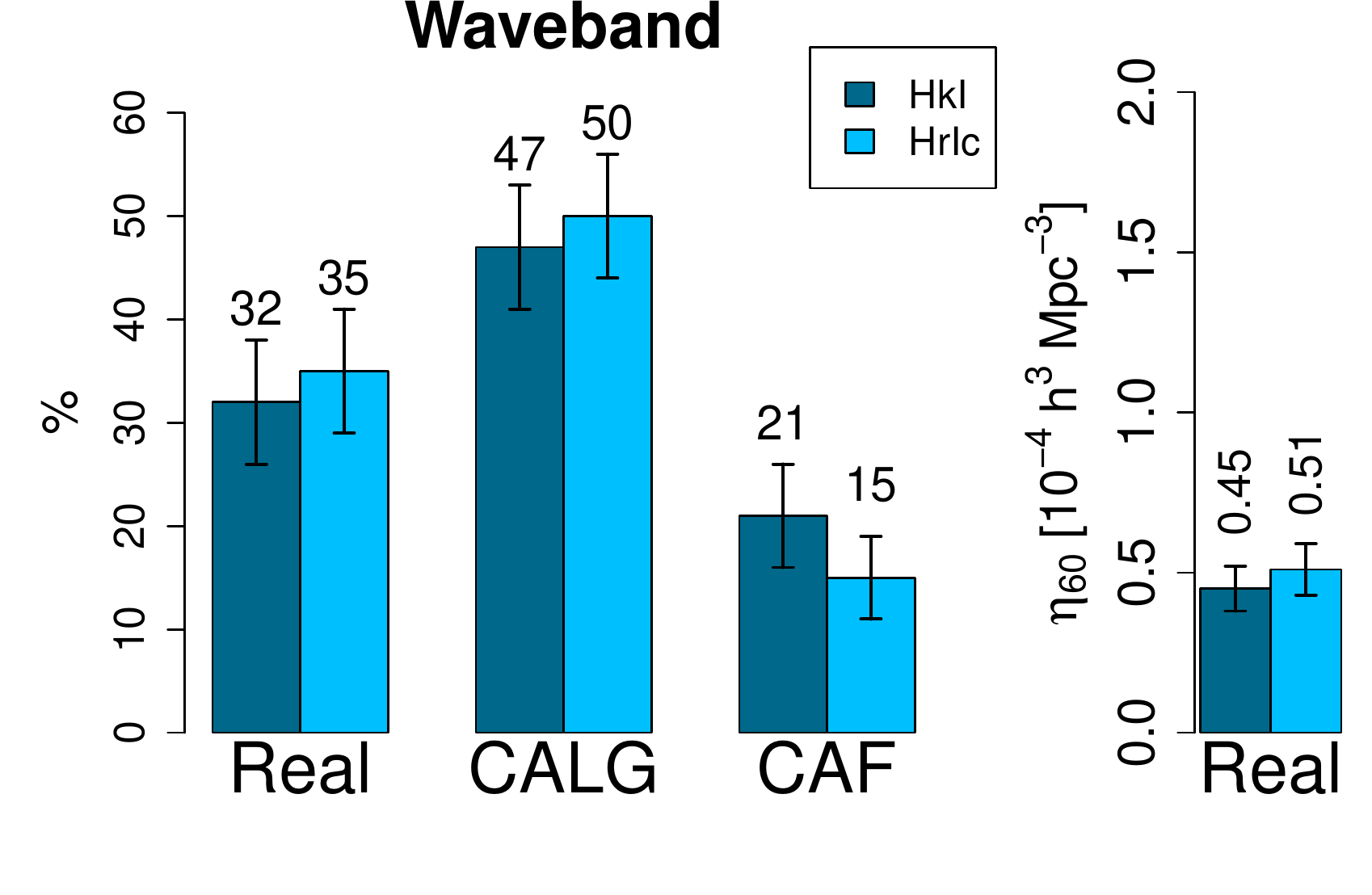}
  \end{minipage}
  \begin{minipage}[b]{0.44\textwidth}
    \centering
    \includegraphics[width=1.0\textwidth,trim=0 170 15 15,clip]{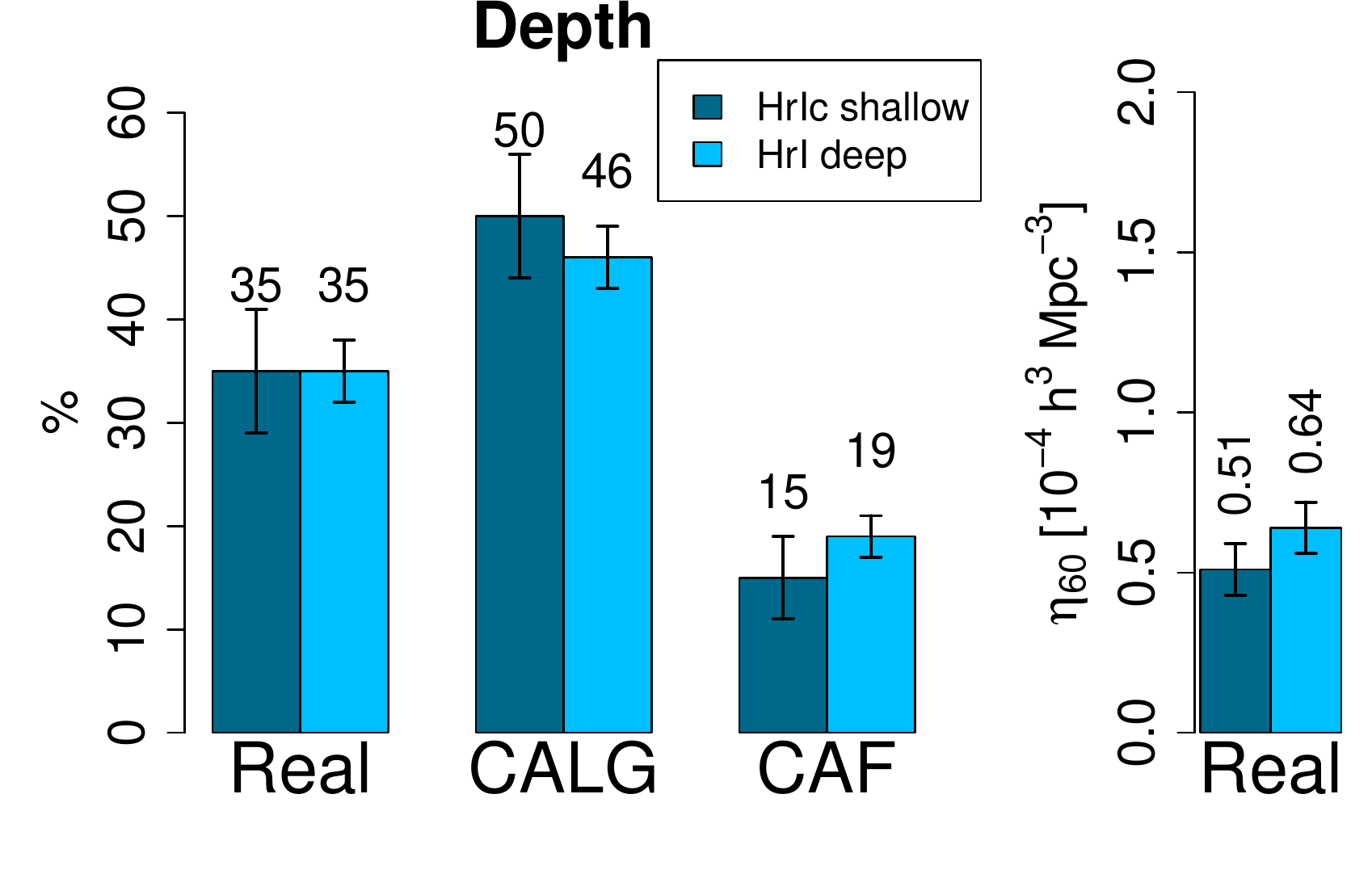}
  \end{minipage}
  \begin{minipage}[b]{0.44\textwidth}
    \centering
    \includegraphics[width=1.0\textwidth,trim=0 170 15 15,clip]{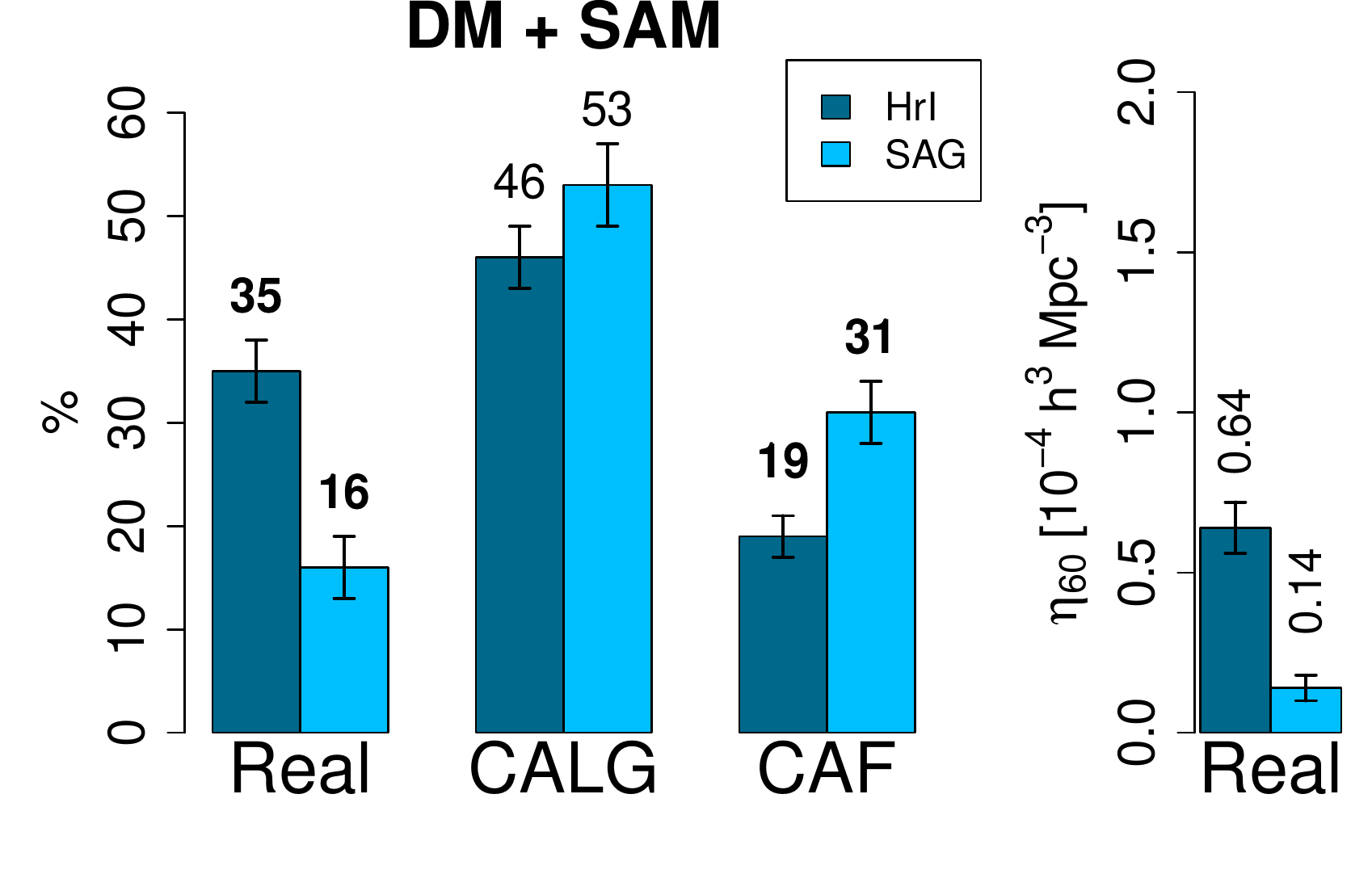}
  \end{minipage}
  \begin{minipage}[b]{0.44\textwidth}
    \centering
    \includegraphics[width=1.0\textwidth,trim=0 170 15 15,clip]{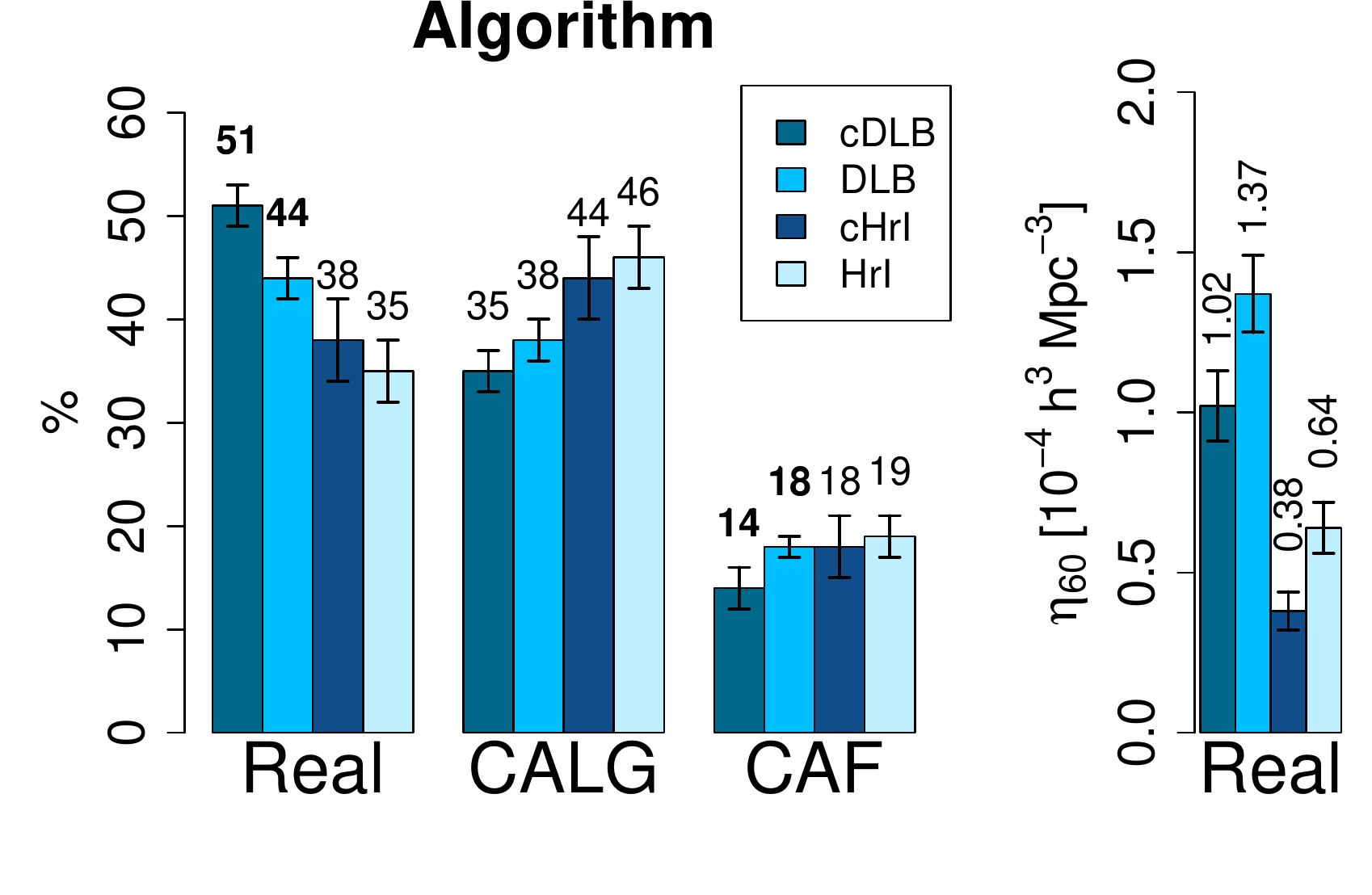}
  \end{minipage}
  \begin{minipage}[b]{0.44\textwidth}
    \centering
    \includegraphics[width=1.0\textwidth,trim=0 170 15 15,clip]{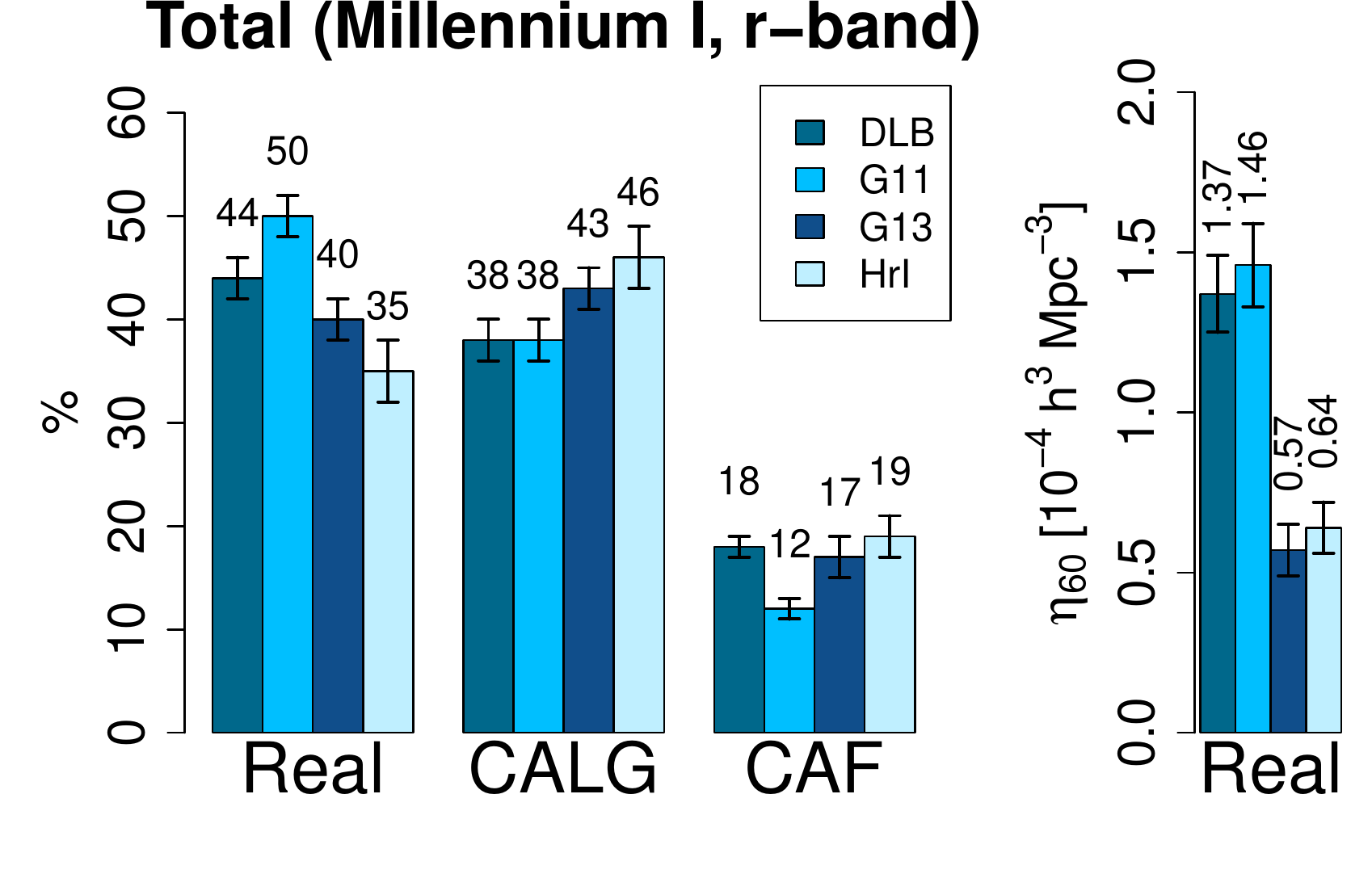}
  \end{minipage}
\caption{Fraction of CGs classified as physically dense (Real), chance
    alignments in loose groups (CALG) or chance alignment in field or
    filaments (CAF). Each panel shows a different comparison as a function of
    cosmology, semi-analytical model (SAM), simulation resolution, apparent
    magnitude selection (waveband), or apparent magnitude limit (depth). We
    also show a total comparison for all the samples obtained from the
    Millennium~I simulation in the $r$ band with $r_{\rm lim}=17.77$. Error
    bars show the 95\% binomial confidence interval for each percentage
    computed as $\pm 1.96 \sqrt{f(1-f)/N_{\rm CG}}$, where $f$ is the fraction. Percentage values in boldface indicate statistically different results taking into account confidence intervals (for all the panels except for the Total Millennium comparison - bottom right panel). 
    }
\label{fbar}
\end{figure*}

The top panel of Figure~\ref{fig:stackbar} displays, 
for the different SAM lightcones, the
percentages of CGs that have been classified as Reals, CALGs and CAFs. The
percentage of CGs classified as Reals ranges from $16\%$ to $56\%$ depending
on the sample that is being analysed, the samples built from the simulation
with the highest resolution (GII and HrII) present the highest fractions of
Reals, while the sample extracted from SAG, which has the poorest early spatial 
resolution (by 0.4 dex) as well as poorest mass resolution (by 0.2 dex) presents the lowest fraction of Real CGs.  Among the chance
alignments, in all the lightcones the percentage of CALGs is higher than the
percentage of CAFs. The chance alignments in the field increase with increase
cosmological density parameter as
\begin{equation}
  f_{\rm CAF} \propto \Omega_{\rm m}^\delta \ ,
\end{equation}
with
\begin{equation}
    \delta = \left\{
    \begin{array}{ll}
    4.1\pm1.7 & ({\rm G11} \to {\rm G13}) \\
    1.1\pm0.3 & ({\rm all \ MS})\\
    \end{array} \right.
    \ ,
\end{equation}
i.e. a stronger dependence than the fraction of physically dense groups with
$\Omega_{\rm m}$ (eq.~[\ref{frealvsOmegam}]), highlighting how the increased density of the Universe lead
to more frequent chance alignments of galaxies in the field.
The fit using all MS-based samples yields a slightly weaker dependence with $\Omega_{\rm m}$ of $f_{\rm CAF}$ in comparison to $f_{\rm Real}$.

\subsection{Effects of different frequency of orphan galaxies}
As discussed in Sect.~\ref{sec:simulations},  differences in the treatment of orphan galaxies in the SAMs impact on the clustering of galaxies at small scales, which are actually the scales of interest for CGs. Therefore, we examined how the fraction of orphan galaxies in the lightcones affects the occurrence of CGs. The right panels in Fig.~\ref{fig:fcosmo} show the space density of CGs (top), the space density of Real CGs (middle) and the fraction of Real CGs (bottom) as a function of the fraction of orphan galaxies in their parent lightcones. 
The percentage of orphans in the lightcones varies from $\sim 2$ to $\sim 9 \%$. 
The top panel of Fig~\ref{fig:fcosmo} indicates that the higher the fraction of orphan galaxies, the higher the space density of CGs. However, it could be tricky to mix all the samples since the frequency of orphan galaxies in the SAMs is not only determined by the particular recipes, but also linked to the resolution of the DM simulation they used to follow the merger trees.  
Considering only those samples extracted from the MS-based lightcones, the Spearman rank correlation coefficients (inserted in Fig.~\ref{fig:fcosmo}) indicate strong correlations between the three properties analysed and the fraction of orphans in the lightcones: higher fractions of orphans lead to higher space densities of CGs and of  Real CGs,  and, especially, a higher fraction of Real CGs. 

Regarding the constitution of the CGs, in the bottom panel of Fig.~\ref{fig:stackbar}, we show the percentages of galaxies of different types that populate the CGs. Central galaxies (type 0) are almost a quarter of the galaxies in CGs in all the lightcones. The percentage of orphan galaxies (type 2) varies from $13$ to $52\%$ depending on the SAM and on the resolution of the parent simulation. 
Even more, if we only consider galaxies in Real CGs, the fraction of orphan galaxies is higher than in the total samples (ranging from 19 to 61\%), which indicates the importance of orphan galaxies in very dense environments. 
The excess of Real CGs will lead to an excess of total number of CGs with a smaller correlation (as shown in Fig.~\ref{fig:fcosmo}) and also to an excess of the fraction of Real CGs (as seen in Fig.~\ref{fig:fcosmo}).

What causes the correlations with the fractions of orphans (Fig.~\ref{fig:fcosmo})? 
Orphans have uncertain positions, since they lost their subhaloes and their positions are guessed using crude recipes for their orbital evolution including dynamical friction. 
Some SAMs will make the orphans live too long, while others will make them disappear too quickly. 
In the first case, we should see a surplus of galaxies in the central regions where the subhaloes should be more affected by tidal forces and tidal heating and destroyed or reduced to less than the allowed minimum number of particles. 
This surplus of galaxies in the inner regions of haloes would lead to more CGs (when isolated), which may explain the positive correlations between CG space density and fraction of orphans. 
\cite{pujol17} show that SAG orphans tend to lie in the outskirts of clusters, while HrI orphans shows no trend with radius.

It is instructive to compare the orphan fractions in similar SAMs at different resolutions.
Figure~\ref{fig:fcosmo} shows that the fractions of orphans in the lightcone decreases by typically 40 per cent when moving from low (G11 and Hr1) to high resolution (respectively GII and HrII) SAMs, while the corresponding fraction of Real CGs increases slightly (G11 to GII) or by a factor 1.5 (HrI to HrII). This leads to a shift of these two high-resolution SAMs from the tight relation between fraction of Reals and orphan fraction seen for the MS-I based SAMs (see arrows in the bottom right panel of Fig.~\ref{fig:fcosmo}).
The fraction of Real CGs appears less dependent of the orphan fraction at higher resolution. This may be due to the fact that the relative flux limits of CGs prevents them from containing the lowest mass galaxies (hence lowest-mass subhaloes) in the MS-II based SAMs, thus these CGs have fewer orphans and the fraction of Reals depends less on the fraction of orphans.

\subsection{Effects of the other parameters}

Given that the samples differ in a number of features, in Fig.~\ref{fbar}, we show the comparison of the percentage of CGs classified into Reals, CALGs and CAFs between pairs of samples that share common features but differ in one at a time, which could be cosmology, semi-analytical modelling, DM resolution, wavebands, depth or identification algorithm.  
A detailed comparison yielded the following results:
\begin{itemize}
\itemsep 0.25\baselineskip

\item to compare the effects of the cosmological parameters on the formation
  of CGs, we compare G11 and G13 ($=$ SAM, $\ne$ cosmology, $=$ mass
  resolution, $=$ depth, $=$ waveband, $=$ finder algorithm): the smaller the
  matter density ($\Omega_{\rm m}$), the higher the fraction of Reals, and
  the smaller the fraction of CAFs, while the fraction of CALGs remains
  almost unchanged.

\item to compare the effects of the SAMs, we use G11 and DLB ($\ne$ SAM, $=$
  cosmology, $=$ mass resolution, $=$ depth, $=$ waveband, $=$ finder
  algorithm): G11 model is more efficient to find Real CGs than DLB. This
  decrement in the fraction of Real systems in DLB is translated in an
  increment of the fraction of CAF systems.

\item to understand the effects of space and mass resolution on the CG
  frequency, we compare G11 and GII, or HrI and HrII ($=$ SAM, $=$ cosmology,
  $\ne$ mass resolution, $=$ depth, $=$ waveband, $=$ finder algorithm): the
  higher the mass resolution of the simulation, the higher the fraction of
  Real CGs and the lower the fraction of CALGs. 
  However, as we noted in the previous section, this correlation might depend strongly on the treatment of orphan galaxies performed in each SAM (for instance, the difference between the fraction of Reals in MS vs MSII is smaller in the \citeauthor{Guo+11} SAM than in \citeauthor{Henriques+15}).

\item to evaluate the effects of the photometric waveband, we analyse HrI and
  HkI ($=$ SAM, $=$ cosmology, $=$ mass resolution, $=$ depth, $\ne$
  waveband, $=$ finder algorithm): there are not significant differences in
  the fraction of Real, CALG or CAF CGs identified in different bands.

\item to understand the effects of the catalogue depths, we use HrI and HrIc
  ($=$ SAM, $=$ cosmology, $=$ mass resolution, $\ne$ depth, $=$ waveband,
  $=$ finder algorithm): there are no significant differences in the fraction
  of CGs of any type when using lightcones restricted to different flux
  limits.

\item to compare different N-body simulation and galaxy formation recipes
  with an unique cosmology, we compare HrI with SAG ($\ne$ SAM, $\ne$ DM
  simulation, $=$ cosmology, $=$ depth, $=$ waveband, $=$ finder algorithm):
  SAG produces a smaller fraction of Real CGs (less
  than half than in HrI), while it has higher fraction of CAFs.

\item to understand the effect of using different CG finder, we analyse cDLB
  and DLB, or cHrI and HrI ($=$ SAM, $=$ cosmology, $=$ mass resolution, $=$
  depth, $=$ waveband, $\ne$ finder algorithm): the fraction of Real CGs when
  using the classic or the modified algorithm is dependent on the SAM. For
  DLB, the fraction of Reals is higher when using the classic algorithm
  instead of the modified; however, this difference is not significant when
  comparing cHrI and HrI (it was also demonstrated in \cite{DiazGimenez+18}).
\end{itemize}

\subsection{Comparison with previous work}

\begin{table*}
\caption{Description of the samples and percentage of Real CGs in previous works}
\begin{center}
{\small
\tabcolsep 1.9pt
    \begin{tabular}{ccccccccccccccccccc}
    \hline
    \hline
     & DGM10 && \multicolumn{2}{|c|}{DG+12} && DGZ15 && \multicolumn{2}{|c|}{TDGZ16} && \multicolumn{2}{|c|}{DGZT18} && \multicolumn{5}{|c|}{this work} \\ 
   Catalogue & box && \multicolumn{2}{|c|}{lightcone} && lightconeH+12 && \multicolumn{2}{|c|}{lightconeH+12} && \multicolumn{2}{|c|}{lightcone} && \multicolumn{5}{|c|}{lightcone}\\
   Blending & Shen+03 && \multicolumn{2}{|c|}{Shen+03} && $1.5\times$Lange15 && \multicolumn{2}{|c|}{Lange+15} && \multicolumn{2}{|c|}{Lange+15} && \multicolumn{5}{|c|}{Lange+15}\\
   \cline{12-13}
   \cline{15-19}
    Algorithm & classic && \multicolumn{2}{|c|}{classic} && classic && \multicolumn{2}{|c|}{classic} && classic & modified && classic & \multicolumn{4}{|c|}{modified}\\
    \cline{4-5}
    \cline{9-10}
    \cline{16-19}
    mag. lim & $17.44$($R$) && $16.3$($r$) & $17.77$($r$) &&  $13.57$($K_s$) && $16.54$($r$) & $13.57$($K_s$) && \multicolumn{2}{|c|}{$17.77$($r$)} && $17.77$($r$) && $17.77$($r$) & $16.54$($r$) & $13.57$($K_s$) \\
    \hline
    DLB & 59 && 53 & 58 && -- && \multicolumn{2}{|c|}{--} && \multicolumn{2}{|c|}{--} && 51 && 44 & -- & -- \\
    G11 & -- &&  \multicolumn{2}{|c|}{--} && 54 && 52 & 56 && \multicolumn{2}{|c|}{--} && --&& 50 & -- & -- \\
    GII & -- &&  66 & 69 && -- && \multicolumn{2}{|c|}{--} && \multicolumn{2}{|c|}{--} && -- && 56 & -- & -- \\
    HrI & -- &&  \multicolumn{2}{|c|}{--} && -- && \multicolumn{2}{|c|}{--} && 38 & 35 && 38 && 35 & 35 & 32 \\
    \hline
    \end{tabular}
    }
    \end{center}
    \label{tab:previous}

\noindent\parbox{\hsize}{Notes: The acronyms are
DGM10 \citep{DiazGimenez&Mamon10},
DG+12 \citep{DiazGimenez+12},
DGZ15 \citep{DiazGimenez+15},
TDGZ16 \citep{Taverna+16},
and
DGZT18 \citep{DiazGimenez+18}.
lightconeH+12 stands for the all-sky lightcone built by \cite{Henriques+12} using the G11 SAM;
Galaxy sizes for estimating blending of galaxies along the line of sight were computed following the relations of \cite{Shen+03} or \cite{Lange+15}.}
\end{table*}

Table~\ref{tab:previous} compares our results on fraction of CGs that are Real with the fractions we produced in previous works. 
The results are all consistent, given the differences in the lightcones, blending, wavebands and limiting magnitude, and the algorithm.
\cite{DiazGimenez&Mamon10} (DGM10) and \cite{DiazGimenez+12} (DG+12) used the DLB SAM (and other SAMs not considered in the present work) to predict slightly higher fractions of Real CGs than found here. The slight differences are caused by the wavebands, lightcones, and galaxy blending method. 
Similarly, the fraction of Real CGs found by \cite{DiazGimenez+15} (DGZ15) and \cite{Taverna+16} (TDGZ16) were slightly higher than what we find here, but this is because we only compare here to our modified algorithm, which produces a lower fraction of Real CGs.
Also, the fraction of Real CGs found with the GII SAM was higher in \cite{DiazGimenez+12} (DG+12), but again we are comparing the classic algorithm to the modified one.
Finally, the Real CG fractions given by \cite{DiazGimenez+18} (DGZT18) are the same as here. The only difference between the two studies for the GII SAM is the higher stellar mass limit in the simulation boxes used in the present work. However, this does not introduce any modification in the sample of CGs. 

\section{Summary}
\label{sec:conclusions}

We analysed samples of compact groups of galaxies extracted from lightcones built from the outputs of 3 different cosmological simulations combined with 5 semi-analytical models of galaxy formation, and using different observational selection functions. A total of 11 samples of CGs has been analysed. We performed a comparative study of frequency and nature of the CGs between the different samples. A summary of the main results is as follows:

- All SAMs are able to produce samples of CGs that in general reproduce well the range of properties of observational CGs.

- The space density of CGs varies from SAM to SAM, with a variation of well over 4
between the SAM with the most frequent CGs (DLB) and the least frequent
(SAG).

- The space density of CGs decreases strongly with increasing density of the Universe, because a higher density Universe will lead to more contamination of the CG isolation criterion.

- The space density of physically dense CGs strongly increases with the normalisation of the power spectrum, $\sigma_8$, as expected. 

- The space density of CGs predicted in the SAMs of \cite{Guo+13,
  Henriques+15, Cora+18} is lower than the expected from observations.

- The space density of CGs is strongly correlated to the fraction of orphan galaxies in the SAMs since they have an important impact in the clustering at small scales (see Fig.~\ref{fig:xi}). 

- The frequency and nature of CGs does not depend on the waveband nor on the flux limit of the samples.

- The fraction of CGs that are physically dense (not caused by chance alignments along the line of sight) vary from 16 to 56 per cent, depending on the SAM.

- The fraction of CGs caused by chance alignments of galaxies along the line of sight increases with higher cosmological density parameter (producing more galaxies along the line of sight). The initial studies performed with the low $\Omega_{\rm m}$ Millennium Simulation thus tend to underestimate the importance of chance alignments.

- The fraction of CGs that are physically dense correlates with the fraction of orphan galaxies in the mock lightcones. Different treatments of orphan galaxies in the models has a direct impact on compact group studies.

- The fraction of CGs caused by chance alignments decreases with increasing resolution of the simulation, because  physically dense groups will suffer less from over-merging of their galaxy subhaloes in a better resolved simulation.

- Our study is the first one based on SAMs
to show that physically dense groups might account for less
than half of observed CGs (as originally proposed by \citealp{Mamon86,Mamon87}
and \citealp{Walke&Mamon89}, although this idea may not survive with better
resolved simulations).  

Therefore, the results obtained from semi-analytical models  on the nature and frequency of compact groups of galaxies must be taken with some caution, as they depend on the cosmological parameters and resolution of the parent simulation on which the models were run, and on the galaxy formation model recipes themselves, particularly regarding the treatment of orphan galaxies.

\section*{Acknowledgements}
         {\small
We thank the anonymous referee for constructive comments (in particular for considering the different treatment of orphan galaxies in the different SAMs). Dr. Sof\'ia Cora for useful discussions about the SAG SAM.
We thank the authors of the SAMs for making their models publicly available. 
The Millennium Simulation databases used in this paper and the web application providing online access to them were constructed as part of the activities of the German Astrophysical Virtual Observatory (GAVO).  
The CosmoSim database used in this paper is a service by the Leibniz-Institute for Astrophysics Potsdam (AIP). The authors gratefully acknowledge the Gauss Centre for Supercomputing e.V. (www.gauss-centre.eu) and the Partnership for Advanced Supercomputing in Europe (PRACE, www.prace-ri.eu) for funding the MultiDark simulation project by providing computing time on the GCS Supercomputer SuperMUC at Leibniz Supercomputing Centre (LRZ, www.lrz.de).
This work has been partially supported by Consejo Nacional de Investigaciones Cient\'\i ficas y T\'ecnicas de la Rep\'ublica Argentina (CONICET) and the Secretar\'\i a de Ciencia y Tecnolog\'\i a de la Universidad de C\'ordoba (SeCyT).}

\appendix
\section{Queries to download SAM data}
\label{queries}
The Millennium
Database\footnote{\label{millennium}\url{https://wwwmpa.mpa-garching.mpg.de/millennium/}}
has timeouts of seven minutes to download data. To work around this, one has
to work with multiple queries.
We followed the suggestions given in their
database\footnote{\url{http://gavo.mpa-garching.mpg.de/Millennium/Help/faqmillennium}},
and ran a script with multiple queries for each of the Millennium-based SAMs. In all cases, we included a safe cut in absolute magnitude limit to speed up the downloads.

\begin{itemize}
\item DLB galaxies were downloaded from the Millennium Database using the following query:
\begin{lstlisting}[
           language=SQL,
           breaklines=true,           
           showspaces=false,
           basicstyle=\ttfamily,
           numbers=left,
           numberstyle=\tiny,
           commentstyle=\color{gray}
        ]
SELECT c.galaxyID,c.type,c.x,c.y,c.z,c.velX,c.velY,c.velZ,c.stellarMass,c.bulgeMass,g.u_sdss,g.g_sdss,g.r_sdss,g.i_sdss,g.z_sdss,g.K_2mass,g.J_2mass
FROM MPAGalaxies..DeLucia2006a c, MPAGalaxies..DeLucia2006a_SDSS2MASS g
WHERE g.galaxyID=c.galaxyID AND c.galaxyID between :START*8*1e12 and (:START+:BIN)*8*1e12-1 AND c.snapnum=$sp AND g.r_sdss<-16 AND c.stellarMass>0.07
\end{lstlisting}

In this SAM, \texttt{:START} ranges from 0-63 and \texttt{:BIN=1} returns 64 separate queries; and \texttt{\$sp} moves from $63$ ($z=0$) to $48$ ($z\sim 0.5$). 

\item The query to download G11 galaxies was:
\begin{lstlisting}[
           language=SQL,
           breaklines=true,           
           showspaces=false,
           basicstyle=\ttfamily,
           numbers=left,
           numberstyle=\tiny,
           commentstyle=\color{gray}
        ]
SELECT galaxyID,type,x,y,z,velX,velY,velZ,stellarMass,bulgeMass,uDust,gDust,rDust,iDust,zDust 
FROM Guo2010a..MR 
WHERE galaxyID between :START*1e12 and (:START+:BIN)*1e12-1 AND snapnum=$sp and rDust<-16 AND stellarMass>0.07
\end{lstlisting}
where \texttt{:START} ranges from 0-511 and \texttt{:BIN=1}; and \texttt{\$sp} moves from $63$ ($z=0$) to $48$ ($z\sim 0.5$). 

\item The query to download GII galaxies was:
\begin{lstlisting}[
           language=SQL,
           breaklines=true,           
           showspaces=false,
           basicstyle=\ttfamily,
           numbers=left,
           numberstyle=\tiny,
           commentstyle=\color{gray}
        ]
SELECT galaxyID,type,x,y,z,velX,velY,velZ,stellarMass,bulgeMass,uDust,gDust,rDust,iDust,zDust 
FROM Guo2010a..MRII 
WHERE galaxyID between :START*16*1e15 and (:START+:BIN)*16*1e15-1 AND snapnum=$sp AND rDust<-11 AND stellarMass>0.007
\end{lstlisting}
where \texttt{:START} ranges from 0-31 and \texttt{:BIN=1}; and \texttt{\$sp} moves from $67$ ($z=0$) to $52$ ($z\sim 0.5$).

\item The query to download G13 galaxies was:
\begin{lstlisting}[
           language=SQL,
           breaklines=true,           
           showspaces=false,
           basicstyle=\ttfamily,
           numbers=left,
           numberstyle=\tiny,
           commentstyle=\color{gray}
        ]
SELECT galaxyID,type,x,y,z,velX,velY,velZ,stellarMass,bulgeMass,uDust,gDust,rDust,iDust,zDust 
FROM Guo2013a..MR7
WHERE galaxyID between :START*1e12 and (:START+:BIN)*1e12-1 AND snapnum=$sp AND rDust<-16 AND stellarMass>0.07
\end{lstlisting}
where \texttt{:START} ranges from 0-511 and \texttt{:BIN=1}; and \texttt{\$sp} moves from $61$ ($z=0$) to $46$ ($z\sim 0.5$). 

\item HrI galaxies are retrieved via the following query:

\begin{lstlisting}[
           language=SQL,
           breaklines=true,           
           showspaces=false,
           basicstyle=\ttfamily,
           numbers=left,
           numberstyle=\tiny,
           commentstyle=\color{gray}
        ]
SELECT galaxyID,type,x,y,z,velX,velY,velZ,stellarMass,bulgeMass,SDSSu_Dust,SDSSg_Dust,SDSSr_Dust,SDSSi_Dust,SDSSz_Dust,Ks_Dust,J_Dust 
FROM Henriques2015a..MRscPlanck1 
WHERE galaxyID between :START*16*1e12 and (:START+BIN)*16*1e12-1 AND snapnum=$sp AND SDSSr_Dust<-16 AND stellarMass>0.07
\end{lstlisting}
where \texttt{:START} ranges from 0-31 and \texttt{:BIN=1}; and \texttt{\$sp} moves from $58$ ($z\sim 0$) to $45$ ($z\sim 0.51$).

\item The query to download HII galaxies was:
\begin{lstlisting}[
           language=SQL,
           breaklines=true,           
           showspaces=false,
           basicstyle=\ttfamily,
           numbers=left,
           numberstyle=\tiny,
           commentstyle=\color{gray}
        ]
SELECT galaxyID,type,x,y,z,velX,velY,velZ,stellarMass,bulgeMass,SDSSu_Dust,SDSSg_Dust,SDSSr_Dust,SDSSi_Dust,SDSSz_Dust,Ks_Dust,J_Dust 
FROM Henriques2015a..MRIIscPlanck1 
WHERE galaxyID between :START*16*1e15 and (:START+BIN)*16*1e15-1 AND snapnum=$sp AND SDSSr_Dust<-11 AND stellarMass>0.007

\end{lstlisting}
where \texttt{:START} ranges from 0-31 and \texttt{:BIN=1}; and \texttt{\$sp} moves from $62$ ($z=0$) to $49$ ($z\sim 0.51$).

\end{itemize}

The Multidark database does not have timeouts and it can be accessed via a Query form in the web browser\footnote{\url{https://www.cosmosim.org/}}. We retrieved data using the following query: 

\begin{lstlisting}[
           language=SQL,
           breaklines=true,           
           showspaces=false,
           basicstyle=\ttfamily,
           numbers=left,
           numberstyle=\tiny,
           commentstyle=\color{gray}
        ]
SELECT GalaxyStaticID,GalaxyType,x,y,z,vx,vy,vz,MstarSpheroid,MstarDisk,MagStarSDSSu,MagStarSDSSg,MagStarSDSSr,MagStarSDSSi,MagStarSDSSz 
FROM `MDPL2`.`SAG` 
WHERE snapnum=$sp AND MagStarSDSSr <= -16 AND (MstarSpheroid+MstarDisk) >= 700000000

\end{lstlisting}
where \texttt{\$sp} moves from $125$ ($z=0$) to $106$ ($z\sim 0.51$).

\section{Orphan galaxies}
\label{sec:orphans}
In each SAM, galaxies are classified as central galaxies of a main halo (type 0), central galaxy of a non-dominant halo, which are satellite galaxies around the central galaxy of the dominant halo (type 1) and orphan galaxies whose subhalo is no longer resolved 
by the simulation (type 2). In general, all galaxies are born as type 0, they usually became type 1 when they fall into a group or cluster and they may became later type 2, which in time merge into the central galaxy of their halo. Each SAM adopts a different treatment of the orphan galaxies (see \citealp{pujol17} for  comparisons of the treatment of orphans in 9 SAMs). 

\begin{figure}
    \centering
    \includegraphics[width=0.95\hsize,viewport=0 30 400 550]{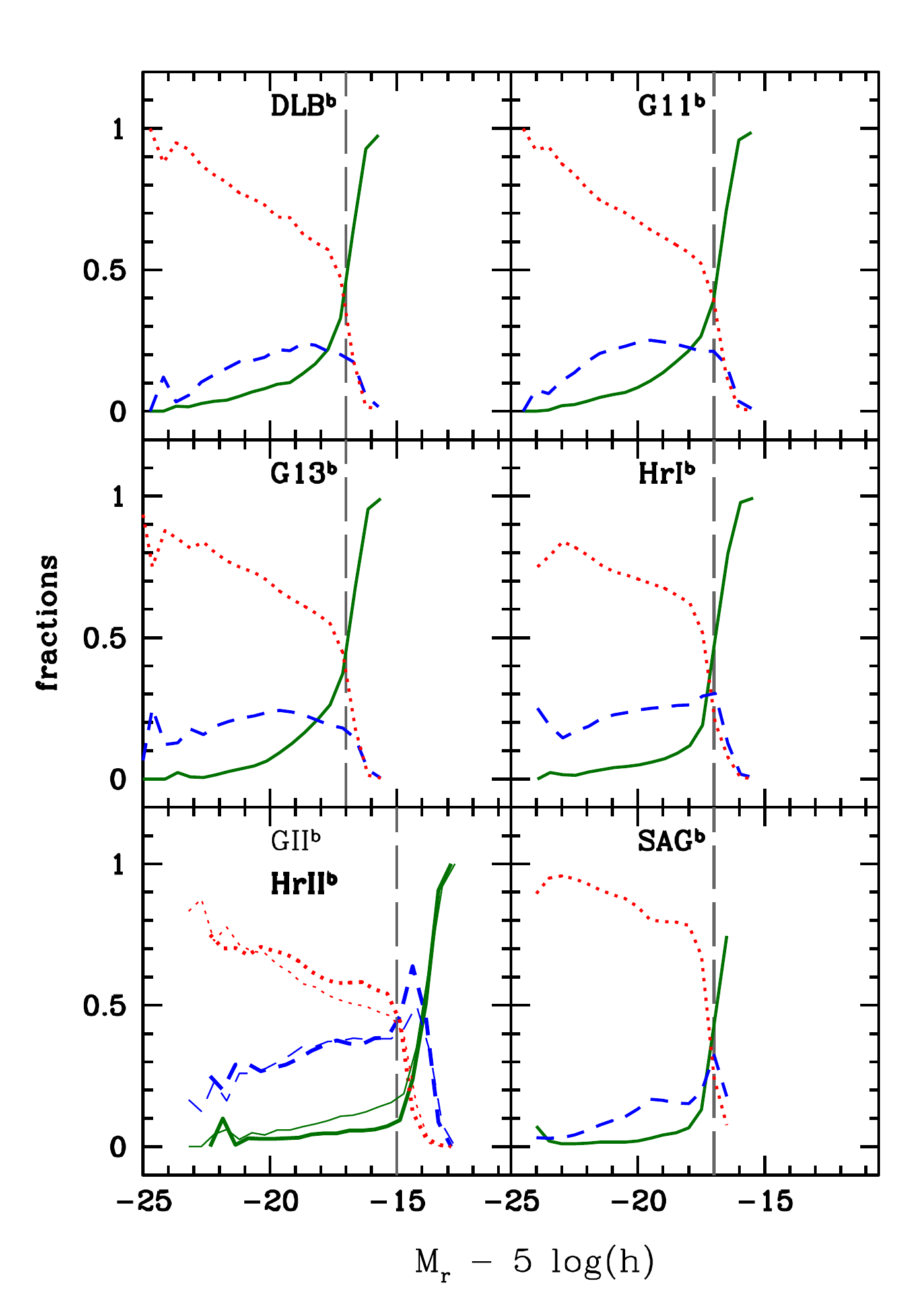}
    \includegraphics[width=0.95\hsize,viewport=0 30 400 570]{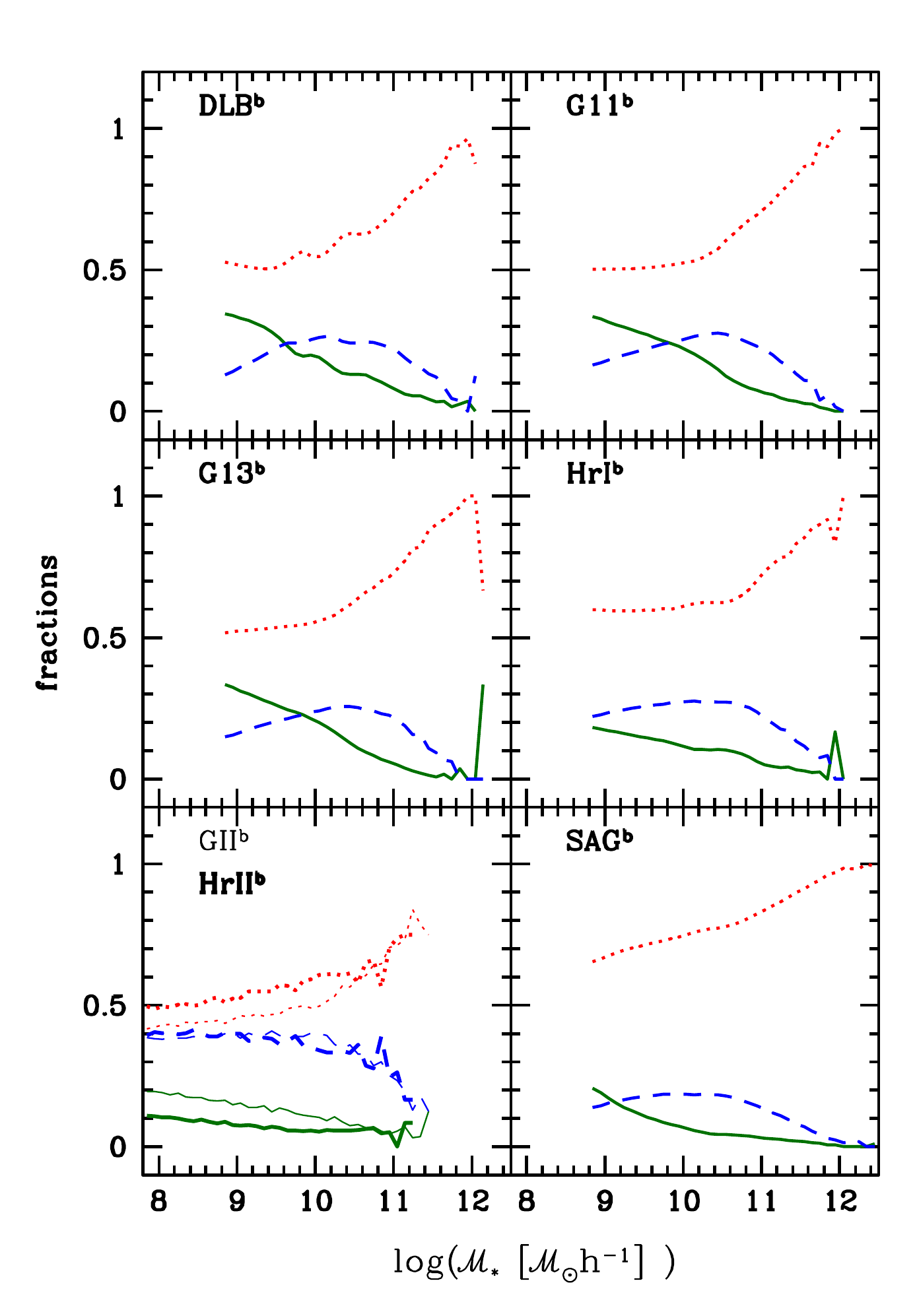}
    \caption{Contributional fraction of galaxies of different types as a function of absolute magnitude (\emph{top panels}) and stellar mass (\emph{bottom panels}). Red dotted lines are central galaxies (type 0), blue dashed lines are satellites (type 1), while green solid lines are orphans (type 2). Vertical lines in the top panels indicate the approximate magnitudes where the samples are quite complete according to Fig.~\ref{fig:fdel}.
    }
    \label{fig:orphan_mag}
\end{figure}

Considering galaxies in the $z$=0 boxes  with stellar masses larger than $7\times10^{8} \, h^{-1} {\cal M}_\odot$ and absolute magnitude $M_r \le -16$, the fraction of orphan galaxies is 0.26, 0.25, 0.25, 0.14 and 0.12 for DLB$^{\rm b}$, G11$^{\rm b}$, G13$^{\rm b}$, HrI$^{\rm b}$ and SAG$^{\rm b}$. 
For the two samples from the MSII ($7\times10^{7} \, h^{-1} {\cal M}_\odot$ and absolute magnitude $M_r \le -11$), the fractions of orphans in the boxes are
lower: they are 0.16 for GII$^{\rm b}$ and 0.09 for HrII$^{\rm b}$.
Figure~\ref{fig:orphan_mag} shows the contributional fraction of galaxies in the simulation boxes as a function of their absolute magnitude (top panels) and stellar masses (bottom panels). Orphan galaxies up to the absolute magnitude where the samples are complete ($\sim -17$) are fewer than $20\%$ in all the SAMs, and they increase towards fainter magnitudes and lower stellar masses. 

\section{Comparison between CG properties from SAMs and observations}
In this section we present a table comparing the physical properties for different CG samples using the well-known Kolmogorov-Smirnov $p$-values and their medians confidence intervals.  

\begin{table*}
\caption{Comparison of physical properties for CGs extracted from SAMs and observations. The comparison is made using Kolmogorov-Smirnov $p$-values, and the $95\%$ confidence intervals for the medians. 
} 
\begin{center}
{\small
\tabcolsep 1.9pt
\begin{tabular}{lccccccccccccccccccc}
    \hline
    \hline
               & \multicolumn{2}{|c|}{$d_{ij}$}  & &  \multicolumn{2}{|c|}{$M_2-M_1$}  & & \multicolumn{2}{|c|}{$\sigma_{\rm v}$}  & & \multicolumn{2}{|c|}{$H_0\,t_{\rm cr}$} & & \multicolumn{2}{|c|}{$r_{\rm p}$} & & \multicolumn{2}{|c|}{$\mu$} \\
\cline{2-3}\cline{5-6}\cline{8-9}\cline{11-12}\cline{14-15}\cline{17-18}               
           & KS-$p$ & M-CI       & & KS-$p$ & M-CI      & & KS-$p$ & M-CI      & & KS-$p$ & M-CI      & & KS-$p$ & M-CI      & & KS-$p$ & M-CI        \\
      \hline     
c2MASS-cDLB  & 0.02    & \checkmark     &&  0.04    & \checkmark && 0.61   & \checkmark && 0.07   & \checkmark && 0.16    & \checkmark     && 0.07    & \checkmark      \\
c2MASS-cHrI & 0.19    & \checkmark     &&  0.51    & \checkmark && 0.30   & \checkmark && 0.14   & \checkmark && 0.35    & \checkmark     && 0.66  & \checkmark      \\
SDSS-DLB    & < 0.01  & \xmark     &&  < 0.01  & \xmark     && < 0.01 & \xmark     && < 0.01 & \xmark     && < 0.01  & \xmark     && 0.10    & \checkmark  \\
SDSS-G11    & < 0.01  & \xmark     &&  < 0.01  & \xmark     && < 0.01 & \xmark     && < 0.01 & \xmark     && < 0.01  & \xmark     && 0.09    & \checkmark  \\
SDSS-GII    & < 0.01  & \xmark     &&  < 0.01  & \xmark     && < 0.01 & \xmark     && < 0.01 & \xmark     && < 0.01  & \xmark     && < 0.01  & \xmark      \\
SDSS-G13    & < 0.01  & \xmark     &&  < 0.01  & \xmark     && < 0.01 & \xmark     && < 0.01 & \xmark     && < 0.01  & \xmark     && 0.77    & \checkmark  \\
SDSS-HrI    & 0.89    & \checkmark &&  0.89    & \checkmark && < 0.01 & \xmark     && < 0.01 & \xmark     && 0.03    & \checkmark && 0.08    & \checkmark  \\
SDSS-HrII   & < 0.01  & \xmark     &&  < 0.01  & \xmark     && < 0.01 & \xmark     && < 0.01 & \xmark     && < 0.01  & \xmark     && 0.95    & \checkmark  \\
SDSS-HkI   & 0.07  & \checkmark     &&  0.70    & \checkmark && < 0.01 & \xmark     && < 0.01 & \xmark     && 0.03  & \checkmark && 0.51  & \checkmark  \\
SDSS-HrIc   & 0.12  & \checkmark     && 0.90  & \checkmark     &&  < 0.01  & \xmark && < 0.01 & \xmark     && 0.15  & \checkmark  && 0.51  & \checkmark  \\
SDSS-SAG    & < 0.01  & \xmark     &&  < 0.01  & \xmark     && < 0.01 & \xmark     && < 0.01 & \xmark     && < 0.01  & \xmark     && < 0.01  & \xmark      \\
    \hline
    \end{tabular}
    }
    \end{center}

\parbox{\hsize}{Notes: First column indicates the selected CG samples to be compared, while the remaining columns shown the selected physical properties that were shown in Fig.~\ref{fig:distr}. For each property, KS-$p$ column indicates the Kolmogorov-Smirnov $p$-value for the given sample comparison, while the M-CI column indicates whether the confidence intervals for the medians of a given property overlaps (\checkmark) or not (\xmark) when comparing the CG samples.
}
    \label{tab:pvalues}
\end{table*}

\bibliography{biblio}
\label{lastpage}
\end{document}